\documentclass[12pt]{article}
\usepackage[latin1]{inputenc}

\usepackage{amsmath}
\usepackage{color}
\usepackage{amsfonts}
\usepackage{amssymb}
\usepackage[mathscr]{euscript}
\usepackage{graphicx}
\usepackage{geometry}
\usepackage{amssymb,epsfig,subfigure}
\usepackage{hyperref}
\usepackage{comment}
\usepackage{tabularx}
\usepackage{bm}
\usepackage{euscript}
\usepackage{graphicx}
\usepackage[dvipsnames]{xcolor}
\usepackage{amsfonts}
\usepackage{exscale}
\usepackage{amsbsy}
\usepackage{subfigure}
\usepackage{textcomp}
\usepackage{comment}
\usepackage{hyperref}
\usepackage{slashed}
\usepackage{authblk}
\usepackage{tabularx}
\usepackage{euscript}
\usepackage{graphicx}
\usepackage{color}
\usepackage{amsfonts}
\usepackage{exscale}
\usepackage{amsbsy}
\usepackage{subfigure}
\usepackage{textcomp}
\usepackage{comment}
\usepackage{hyperref}
\usepackage{bm}
\usepackage{wrapfig}
\usepackage[font=footnotesize,labelfont=bf]{caption}
\usepackage{ulem} % Strike-through package. REMOVE AFTER REVISIONS
\usepackage{makecell}

\def\ba#1\ea{\begin{align}#1\end{align}}
\def\bg#1\eg{\begin{gather}#1\end{gather}}
\def\bm#1\em{\begin{multline}#1\end{multline}}
\def\bmd#1\emd{\begin{multlined}#1\end{multlined}}

\newcommand{\be}{\begin{equation}}
\newcommand{\ee}{\end{equation}}
\newcommand{\bea}{\begin{eqnarray}}
\newcommand{\eea}{\end{eqnarray}}

\newcommand{\pd}{\partial}

\newcommand{\matleft}{\left(\begin{array}}
\newcommand{\matright}{\end{array}\right)}
\newcommand{\Tr}{\operatorname{Tr}}

\usepackage[numbers,sort&compress]{natbib}
\def\simge{%  ``greater than about'' symbol
    \mathrel{\rlap{\raise 0.511ex 
        \hbox{$>$}}{\lower 0.511ex \hbox{$\sim$}}}}

\def\simle{%  ``less than about'' symbol
    \mathrel{\rlap{\raise 0.511ex 
        \hbox{$<$}}{\lower 0.511ex \hbox{$\sim$}}}}

\makeatletter
\renewcommand\section{\@startsection {section}{1}{\z@}%
                                 {-3.5ex \@plus -1ex \@minus -.2ex}%nn
                                   {2.3ex \@plus.2ex}%
                                   {\normalfont\large\bfseries}}
\renewcommand\subsection{\@startsection{subsection}{2}{\z@}%
                                   {-3.25ex\@plus -1ex \@minus -.2ex}%
                                     {1.5ex \@plus .2ex}%
                                     {\normalfont\bfseries}}
\renewcommand\subsubsection{\@startsection{subsubsection}{3}{\z@}%
                                   {-3.25ex\@plus -1ex \@minus -.2ex}%
                                     {1.5ex \@plus .2ex}%
                                     {\normalfont\itshape}}
\makeatother

\def\pplogo{\vbox{\kern-\headheight\kern -29pt
\halign{##&##\hfil\cr&{\ppnumber}\cr\rule{0pt}{2.5ex}&\ppdate\cr}}}
\makeatletter
\def\ps@firstpage{\ps@empty \def\@oddhead{\hss\pplogo}%
  \let\@evenhead\@oddhead % in case an article starts on a left-hand page
}%      The only change in \maketitle is \thispagestyle{firstpage} instead of 
\def\maketitle{\par
 \begingroup
 \def\thefootnote{\fnsymbol{footnote}}
 \def\@makefnmark{\hbox{$^{\@thefnmark}$\hss}}
 \if@twocolumn
 \twocolumn[\@maketitle]
 \else \newpage
 \global\@topnum\z@ \@maketitle \fi\thispagestyle{firstpage}\@thanks
 \endgroup
 \setcounter{footnote}{0}
 \let\maketitle\relax
 \let\@maketitle\relax
 \gdef\@thanks{}\gdef\@author{}\gdef\@title{}\let\thanks\relax}
\makeatother

\hypersetup{
    unicode=false,          % non-Latin characters 
    pdftoolbar=true,        % show Acrobat
    pdfmenubar=true,        % show Acrobat 
    pdffitwindow=false,     % window fit to page when opened
    pdfstartview={FitH},    % fits the width of the page to the window
    pdftitle={Non-Abelian Fermionization and the QH Landscape},    % title
    pdfauthor={},     % author
    pdfsubject={Subject},   % subject of the document
    pdfcreator={},   % creator of the document
    pdfproducer={}, % producer of the document
    pdfkeywords={keyword1} {key2} {key3}, % list of keywords
    pdfnewwindow=true,      % links in new window
    colorlinks=true,       % false: boxed links; true: colored links
    linkcolor=OliveGreen, %red,          % color of internal links (change box color with linkbordercolor)
    citecolor=NavyBlue,        % color of links to bibliography
    filecolor=magneta,      % color of file links
    urlcolor=cyan           % color of external links
}

\numberwithin{equation}{section}

\newcommand*\samethanks[1][\value{footnote}]{\footnotemark}

\newcommand\blfootnote[1]{%
  \begingroup
  \renewcommand\thefootnote{}\footnote{#1}%
  \addtocounter{footnote}{-1}%
  \endgroup
}
\begin{document}

\normalem

\setcounter{page}0
\def\ppnumber{\vbox{\baselineskip14pt
%\hbox{hep-th/0000000}
}}

\def\ppdate{%\footnotesize{SU/ITP-14/XX}
} \date{\today}

%\title{\LARGE \bf Non-Abelian Fermionization and the Read-Rezayi Sequence}
\title{\LARGE \bf Non-Abelian Fermionization and the Landscape of Quantum Hall Phases}
\author{Hart Goldman$^{1,\psi}$}
\author{Ramanjit Sohal$^{2,\chi}$}
\author{Eduardo Fradkin$^2$}
\affil{$^1$ \it \small Department of Physics, Massachusetts Institute of Technology, Cambridge, MA 02139, USA  \vspace{-4mm}}
\affil{$^2$ \it \small Department of Physics and Institute for Condensed Matter Theory, \vspace{-1.5mm} \\ \it  \small University of Illinois at Urbana-Champaign, %\\  \it 1110 West Green Street, 
Urbana, Illinois 61801, USA}
\maketitle

\begin{abstract}
%Using recently proposed non-Abelian boson-fermion dualities in 2+1 dimensions, we construct parent theories of non-Abelian quantum Hall states starting from multilayer Abelian theories. 

The recent proposal of non-Abelian boson-fermion dualities in 2+1 dimensions, which morally relate $U(k)_N$ to $SU(N)_{-k}$ Chern-Simons-matter theories, presents a new platform for exploring the landscape of non-Abelian quantum Hall states accessible from theories of Abelian composite particles. Here we focus on dualities relating theories of Abelian quantum Hall states of bosons or fermions to theories of non-Abelian ``composite fermions'' partially filling Landau levels. We show that these dualities predict special filling fractions where both Abelian and non-Abelian composite fermion theories appear capable of hosting distinct topologically ordered ground states, one Abelian and the other a non-Abelian, $U(k)_2$ Blok-Wen state. %, in apparent conflict with the duality. 
Rather than being in conflict with the duality, we argue that %this contradiction can be resolved by 
these results indicate unexpected dynamics %in the non-Abelian theory, 
in which the infrared and lowest Landau level limits fail to commute across the duality. %, and
In such a scenario, the non-Abelian topological order can be destabilized in favor of the Abelian ground state, suggesting the presence of a phase transition between the Abelian and non-Abelian states that is likely to be first order. We also generalize these constructions to other non-Abelian fermion-fermion dualities, in the process obtaining new derivations of a variety of paired composite fermion phases using duality, including the anti-Pfaffian state.
%We discuss the implications of our observations for parton constructions of non-Abelian states as well as for our broader understanding of the dynamics of non-Abelian Chern-Simons-matter theories.
Finally, we describe how, in multilayer constructions, excitonic pairing of the composite fermions across $N$ layers can also generate the family of Blok-Wen states with $U(k)_2$ topological order.\blfootnote{$^{\psi\,\leftrightarrow\,\chi}$ These authors contributed equally to the development of this work.}  %\textcolor{violet}{We find, more generally, that a sequence of non-Abelian states which can be described as paired states of composite fermions in an Abelian theory have dual descriptions as IQH states of composite fermions in a dual non-Abelian theory.}
\end{abstract}
\bigskip
\newpage

{
\hypersetup{linkcolor=black}
\tableofcontents
}
\thispagestyle{empty}

\newpage
\setcounter{page}{1}

\section{Introduction}
\label{sec:introduction}

\begin{comment}
Since the discovery of the fractional quantum Hall (FQH) effect, a rich tapestry of phases of two-dimensional electron gases in magnetic fields has been uncovered experimentally. An even more diverse array of quantum Hall phases have been predicted theoretically, although often through the use of ideal wave functions that are not tied to any specific microscopic physics. Surprisingly, many of these phases are amenable to field theoretic treatments by way of composite boson or composite fermion constructions. In particular, Abelian FQH states may be understood as arising from the formation of integer quantum Hall (IQH) states of composite fermions, or, equivalently, the condensation of composite bosons. 
In contrast, non-Abelian FQH states have eluded a comprehensive, field theoretic description. Although the composite fermion picture may be used to describe a the non-Abelian Moore-Read states, composite particle pictures have largely been unsuccessful in explaining the larger landscape of conjectured non-Abelian FQH states.
\end{comment}

Two-dimensional electron fluids in strong magnetic fields are capable of hosting a rich tapestry of gapped, incompressible phases. The most famous among them occur at partial Landau level (LL) fillings $\nu$ and have fractionally quantized Hall conductivities $\sigma_{xy}=\nu\,\frac{e^2}{h}$, in what is known as the fractional quantum Hall (FQH) effect. An explanation of the large family of FQH phases observed in experiments thus far, as well as an understanding of their underlying topological orders, was a major achievement of the past several decades.  %\sout{of twentieth century physics.} 
This progress is built on the unifying framework of flux attachment \cite{Wilczek-1982}, which relates the original problem of electrons to a new problem of composite fermions or bosons coupled to a fluctuating Chern-Simons gauge field \cite{Witten1989}. %is provided by many of these phases are amenable to field theoretic treatments by way of composite boson or composite fermion constructions. In particular,
For example, in the language of flux attachment, the observed Abelian FQH states may be understood as arising from the formation of integer quantum Hall (IQH) states of composite fermions \cite{Jain-1989,Lopez-1991}, or, equivalently, the condensation of composite bosons \cite{Zhang-1989,Wen1995}. 

%Besides 
Beyond the experimentally visible FQH states, an even more diverse array of quantum Hall phases have been proposed theoretically, many of which %involve non-Abelian topological orders, which 
host quasiparticles with non-Abelian braiding statistics \cite{Wen1991,Moore1991,Blok1992,Read1999,Ardonne-2004}. %\textcolor{blue}{\bf We need to check what references we need here.} 
These proposals generally lack reference to microscopic physics, instead being based on ``ideal'' wave functions.   % although often through the use of trial (``ideal'') wave functions that are disconnected from any specific microscopic physics. Surprisingly, many of these phases are amenable to field theoretic treatments by way of composite boson or composite fermion constructions. In particular, Abelian FQH states may be understood as arising from the formation of integer quantum Hall (IQH) states of composite fermions, or, equivalently, the condensation of composite bosons. 
Indeed, a general characterization of the dynamics that may lead to such non-Abelian phases has proven elusive, and it is unclear which of these phases are accessible from physically motivated theories of (Abelian) composite particles. %With the exception of  the non-Abelian Moore-Read (MR) states, which can be understood as paired states of composite fermions \cite{Read2000}, 
With some notable exceptions, in which non-Abelian FQH states are obtained as paired states of composite fermions (the Moore-Read state) \cite{Read2000} or composite bosons (the Read-Rezayi sequences) \cite{Fradkin1998,Fradkin1999,Goldman2019},  the composite particle picture %\textcolor{blue}{(except for the case of bosonic Read-Rezayi (RR) states \cite{ Fradkin1999,Goldman2019})}
%, which are inherently Abelian, 
has been largely unsuccessful in accessing the wider landscape of conjectured non-Abelian FQH states. 

In recent work \cite{Goldman2019}, we made progress %towards filling this gap in our understanding 
on mapping the region of the non-Abelian FQH landscape accessible to theories of Abelian composite particles. We developed Landau-Ginzburg theories for a large class of non-Abelian states which are related to Abelian composite particle theories via %. To do so, we
%by employing a set of 
recently proposed Chern-Simons-matter theory dualities \cite{Aharony2016}. Motivated by the equivalence of $U(N)_k$ Chern-Simons theories coupled to gapless complex bosons and $SU(k)_{-N}$ theories coupled to gapless Dirac fermions in the 't Hooft (large-$N,k$) limit \cite{Aharony2012,Giombi2012}, these dualities relate theories of gapless bosons or fermions coupled to Chern-Simons gauge theories in a manner which parallels the established level-rank dualities of pure Chern-Simons theories \cite{Naculich-1990,Naculich-1990b,Camperi-1990,Hsin2016}. Of these dualities, several relate theories with Abelian and non-Abelian gauge groups, meaning that they represent dualities between the conventional composite boson Landau-Ginzburg theories for certain Abelian FQH states and theories of dual bosons coupled to non-Abelian Chern-Simons gauge fields.\footnote{We emphasize here that it is possible for a Chern-Simons gauge theory to have a non-Abelian gauge group but an Abelian braid group, meaning that it represents an Abelian topological phase. For example, $SU(2)_{1}$ is Abelian in this sense, having the same anyon content as $U(1)_{2}$ by level-rank duality.} By stacking multiple copies of these Abelian states and introducing pairing in the dual non-Abelian composite boson language, we showed that one can access the Read-Rezayi \cite{Read1999} and generalized non-Abelian spin singlet states \cite{Ardonne1999,Ardonne2001,Fuji2017}. The success of this construction stems from the observation that phases naturally accessible by condensing local operators in the non-Abelian dual theory may not be visible in the original Abelian theory, in which these operators correspond to non-local monopole operators. %\textcolor{blue}{\cite{Seiberg2016}}
\begin{comment}
This effective field Theory B$'$escription nicely mirrors the formulation of the %ideal 
wave functions for these states, which are obtained %through the action of 
via a projection %on
from multi-layer Abelian wave functions.
\end{comment}

%Now, 
Lying at the heart of our construction is a %for the Read-Rezayi states was a %specific 
string of dualities involving the usual Landau-Ginzburg theory for the $\nu=1/2$ bosonic Laughlin state, a single flavor of Wilson-Fisher boson (describing the Laughlin quasiparticles) coupled to a $U(1)_2$ Chern-Simons gauge field. This theory has three duals, 
\begin{align}
\text{a Wilson-Fisher scalar}+U(1)_{2}&\longleftrightarrow\text{a Dirac fermion}+SU(2)_{-1/2}\nonumber\\
\label{eq: SU(2) dualities, intro}
\updownarrow\hspace{1.2cm}&\\
\text{a Dirac fermion}+U(1)_{-3/2}&\longleftrightarrow\text{a Wilson-Fisher scalar}+SU(2)_1\,,\nonumber
\end{align}
where we use $\longleftrightarrow$ to denote duality and subscripts denote the Chern-Simons level (including the parity anomaly). The Abelian boson-fermion duality featured here and others like it were explored in Refs. \cite{Seiberg2016,Karch2016,Mross2017,Goldman2018}. While most non-Abelian dualities are boson-fermion dualities, this \emph{quadrality} -- in which each of the four theories is dual to the others -- %\textcolor{blue}{(i.e. a duality between pairs of mutually dual theories)}
 is distinguished by its inclusion of non-Abelian boson-boson and fermion-fermion dualities. %, a consequence of the level-rank duality, $U(1)_2\leftrightarrow U(1)_{-2}$ \cite{Hsin2016}. 
 In Ref. \cite{Goldman2019},  we focused on the non-Abelian boson-boson duality, in which the dual theory consists of non-Abelian bosonic ``composite vortices'' coupled to a $SU(2)_1$ Chern-Simons gauge field, obtaining the  non-Abelian phases via inter-layer pairing of the composite vortices. 

 %Although the construction of non-Abelian states 
%is %most 
%quite transparent in this bosonic description, %these theories are also dual to a theory of fermions coupled to $U(1)_{-3/2}$ Chern-Simons and a theory of fermions coupled to $SU(2)_{-1/2}$ Chern-Simons; 
In this work, we study the non-Abelian phases accessible to the dual theories of \emph{composite fermions}. Of these, the theory of Dirac fermions coupled to a $U(1)_{-3/2}$ gauge field is a relativistic version of the standard composite fermion description of the $\nu=1/2$ bosonic Laughlin state, while the theory of Dirac fermions coupled to a $SU(2)_{-1/2}$ gauge field constitutes a different kind of ``flux attachment'' in which the composite fermions possess charge under a fluctuating non-Abelian gauge field. %Abelian \textcolor{blue}{(with gauge group $U(1)$)} and non-Abelian \textcolor{blue}{theories (with gauge group $SU(2)$) of} ``composite fermions.'' %we will refer to these as Abelian and non-Abelian composite fermion theories, respectively. 
%We left the question of how our construction could be interpreted in this fermionic language unanswered, which partly serves as the motivation for the present paper. 
%Indeed, 
%In particular, 
%Indeed, in this work, we will show that a stacking and pairing procedure analogous to the one described in Ref. \cite{Goldman2019} can be used to obtain the (time-reversed conjugates of) the Read-Rezayi sequence as excitonic phases of the $SU(2)$ fermions. %\sout{ the non-Abelian} composite fermions \textcolor{blue}{of the non-Abelian theory}. 
%We also describe the types of phases accessible from condensing higher-body operators, as in quartetting.
%Such phases can be shown to have enriched  anyon spectra involving additional Majorana fermions. %\textcolor{blue}{\bf (do we need a reference for excitonic pairing?)} {\hart (I don't think so, although I guess we can cite Jackiw and Rossi or something here if we think a reference is necessary.)} % through a stacking and pairing procedure analogous to that employed in the composite boson description. 
Using this duality, we analyze two of the simplest paths to non-Abelian phases:
\pagebreak
\begin{enumerate}
\item Forming integer quantum Hall (IQH) phases of the $SU(2)$ composite fermions, in analogy with Jain's construction of Abelian FQH phases as IQH states of composite fermions \cite{Jain-1989} and earlier projective parton constructions of non-Abelian FQH states \cite{Wen1991,Wen1999,Barkeshli2010}.
\item Excitonic pairing between layers of $SU(2)$ composite fermions. This construction is a composite fermion version of the one presented in Ref. \cite{Goldman2019} for composite bosons.
\end{enumerate}
In both constructions, we will find that the composite fermions yield the $\nu=k/2$ Blok-Wen states with $U(k)_2$ topological order \cite{Blok1992}, in contrast to the Read-Rezayi states obtained via inter-layer pairing of the $SU(2)$ bosons in Ref. \cite{Goldman2019}. For many of these states, these are the first constructions starting from parent theories of Abelian composite particles, rather than projective parton constructions \cite{Wen1999,Barkeshli2010,Barkeshli2010a} or more general non-Abelian/non-Abelian dualities \cite{Radicevic2016}. In addition, by considering more general non-Abelian dualities, we find not only the exotic Fibonacci state \cite{Nayak2008}, but also composite fermion descriptions of a variety of non-Abelian states that have previously been understood via pairing instabilities of a composite Fermi liquid, including a new description of the anti-Pfaffian state \cite{Lee2007,Levin2007}. Remarkably, we find in these special cases that an IQH phase of the non-Abelian composite fermion theory is \emph{dual} to pairing in the usual Abelian description. Our construction of all of these states using non-Abelian dualities represents the first category of main results of this work.

%However, our primary focus in the present work will instead be to reveal an intriguing subtlety of this particular set of dualities which we encountered in constructing said non-Abelian states and, to our knowledge, has not yet been addressed in the literature. 
%Indeed, 
%Although excitonic pairing is a natural route to non-Abelian phases, %one might expect that there are simpler ways of %generating
%forming quantum Hall phases of fermions, as they satisfy the Pauli exclusion principle. 
%our primary focus in this work will be on an apparently simpler means of generating quantum Hall phases from non-Abelian theories of composite fermions. 
In addition to revealing paths to different non-Abelian phases, these non-Abelian fermion-fermion dualities possess surprising information about the dynamics of composite fermions, leading to our second main family of results. These results relate to our first construction of the $U(k)_2$ states, in which a magnetic field and chemical potential are adjusted so that the $SU(2)$ composite fermions fill $k$ Landau levels. %By adjusting an external magnetic field and chemical potential, %the Abelian 
%composite fermions can form IQH phases by filling LLs, corresponding to non-trivial topological orders of the underlying charged degrees of freedom. 
%When applied to Abelian theories of %\sout{Abelian} 
%composite fermions, %theory above, 
%this principle leads to the Jain sequences of Abelian FQH states, and it is natural to expect the same ideas to hold for non-Abelian theories as well. %of Abelian bosonic FQH states. % focusing on the dual fermionic theories,
%Furthermore, one might na\"{i}vely expect the the same physics to hold for the%should also be true of the 
%non-Abelian composite fermions -- 
%Indeed, this intuition is what guided earlier constructions of non-Abelian FQH states using projective parton constructions \cite{Wen1991,Wen1999,Barkeshli2010}. This would indicate that filling $k$ LLs of the %\sout{non-Abelian} \textcolor{blue}{doublet of composite fermions that carry the spin 1/2 representation of $SU(2)$} 
%$SU(2)$ (doublet) composite fermions should 
This leads to a $SU(2)_{-k}$ spin topological quantum field theory (spin TQFT) at low energies, corresponding to $U(k)_2$ topological order by level-rank duality. % as well. 
%will show that when one turns on a magnetic field, 
However, this conclusion is immediately complicated by the duality with the Abelian composite fermion theory, IQH phases of which correspond to the bosonic Jain sequence states. Indeed, there are certain  filling fractions $\nu_*$ of the underlying electric charges at which \emph{both} %the %\textcolor{blue}{composite fermions of Abelian and non-Abelian theories} \sout{composite fermions} 
types of composite fermions fill up an integer number of Landau levels. On integrating out the fermions in these two theories, one would then be led to conclude that the theory with an Abelian gauge group predicts %former predicts 
an Abelian FQH state, while the %the latter 
non-Abelian theory predicts a non-Abelian FQH state. For instance, for a system of bosons at filling $\nu_* = 3/2$, the two theories appear to respectively predict $U(1)_{-2}$ and $U(3)_{2}$ topological order.  

%\textcolor{blue}{\bf EF: The terminology of ``non-Abelian" (or Abelian for that matter) composite particles is imprecise and can lead to confusion: what is Abelian or non-Abelian is the theory, not the fermions themselves (composite or otherwise)!. It is better to state that in once theory the fermions carry the fundamental (spin 1/2) representation of $SU(2)$ and in the other the fundamental charge of $U(1)$. I made changes in the text to this effect.}

\begin{figure}
\centering
\includegraphics[width=0.65\textwidth]{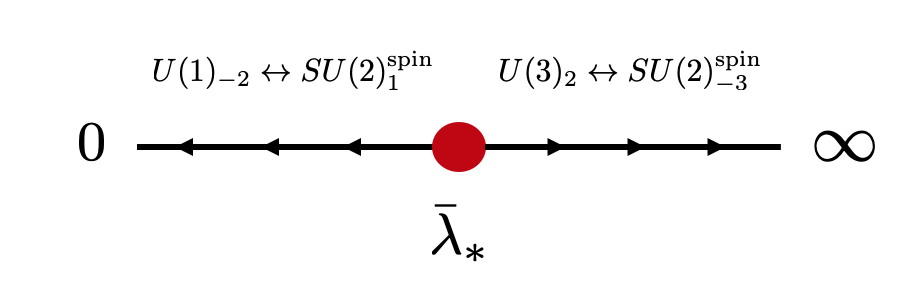}
\caption{Proposed phase diagram for the $SU(2)$ composite fermion theory at filling $\nu=3/2$. Here $\bar\lambda=g_{YM}^2/\omega_c$, where $g_{YM}^2$ is the Yang-Mills coupling (the fine structure constant) and $\omega_c\sim\sqrt{B}$ is the cyclotron frequency. When $\bar\lambda\rightarrow\infty$, the Yang-Mills term vanishes, and the picture of deconfined composite fermions filling (color degenerate) Landau levels is valid, leading to the $SU(2)_{-3}^{\mathrm{spin}}\leftrightarrow U(3)_2$ state. %(the superscript is meant to emphasize that the TQFT is spin, i.e. couples to fermionic matter). 
When $\bar\lambda$ runs small, the Yang-Mills term becomes very large, Landau levels can mix, and the deconfinement of the composite fermions is no longer assured, leading ultimately to the $U(1)_2$ phase predicted in the dual Abelian theory. These states are separated by a critical point at $\bar\lambda=\bar\lambda_*\sim\mathcal{O}(1)$, which is likely first order.} 
\label{fig: SU(2)3 phase diagram}
\end{figure}

%Given the number of checks that Aharony's dualities and their descendents have passed in recent years, it is unlikely that such results signal their breakdown. 
Because $U(3)_{2}$ and $U(1)_{-2}$ are distinct topological orders and are certainly not dual to one another, one might na\"{i}vely worry that these results signal a breakdown of the dualities, which postulate an equivalence of the infrared (IR) limits, or ground states, of the dual theories. % is unlikely that such results signal the breakdown of the dualities, which have survived a number of checks in recent years. 
%However, 
On the other hand, it is very common for %\textcolor{blue}{states with} 
states with distinct topological orders to %\sout{correspond to} \textcolor{blue}{exist at} 
exist at the same filling fraction, with the ultimate choice of ground state depending on %\sout{ultraviolet (UV)} 
details of local energetics. Indeed, in this work we take the view that both topological orders are valid ground states at filling $\nu_*$, %but the dueling this conflict implies
and that the particular choice of ground state depends on the order in which the lowest Landau level ($B\rightarrow\infty$) and IR limits are taken. This order of limits is subtle, as the duality is only valid in the IR limit, while the statement that the composite fermions form a stable IQH state relies on the $B\rightarrow\infty$ limit. %in which duality is implying unexpected dynamics in the non-Abelian theory. %, and we will present a scenario for which the dualities remain intact. 
%Loosely speaking, 
%Such a scenario is possible if the lowest Landau level (LLL) and IR limits fail to commute. 
More precisely, on tuning the ratio of the Yang-Mills coupling to the cyclotron frequency in the %\sout{non-Abelian} composite fermion \textcolor{blue}{of the non-Abelian gauge theory}
non-Abelian theory, we argue that there is a phase transition between the Abelian and non-Abelian FQH states (see Figure \ref{fig: SU(2)3 phase diagram}).  Such a transition, if continuous, would be quite exotic, as it would separate two very different topological orders and therefore lie beyond the Landau-Ginzburg paradigm. Thus, when the $U(1)$ %\sout{Abelian} 
composite fermions %of the Abelian theory 
form an IQH state, the %\sout{non-Abelian} 
$SU(2)$ composite fermions %\textcolor{blue}{of the non-Abelian theory} 
experience an instability and find themselves on the \emph{Abelian} side of this transition and vice versa. %We further illustrate that the transition from the non-Abelian phase to the Abelian phase can potentially arise from the condensation of a composite object of \sout{the non-Abelian} composite fermions \textcolor{blue}{of the non-Abelian theory}. 

Even if this transition is first order, our results demonstrate that QFT dualities can be used to infer non-trivial statements about the phase diagrams and dynamics of Chern-Simons-matter theories in the presence of background fields. Indeed, the phenomenon described above is a general feature of fermion-fermion dualities involving non-Abelian gauge groups, and we collect several additional examples. These appear both in the $SU(2)$ quadrality discussed above as well as in other dualities. %We note that although our focus is on the the dual theories present in the above $SU(2)$ quadrality, our considerations apply to other dualities relating Abelian and non-Abelian Chern-Simons-fermion theories, suggesting that they may in fact be quite general. 
In particular, we consider the case of the duality between a free Dirac fermion and a Dirac fermion coupled to a $U(N)_{-1/2}$ gauge field. We also comment that the scenario we present is reminiscent of recent proposals for the phase diagram of $SU(N)_k$ Chern-Simons theory with $N_f>2k$ fermion flavors (at zero density and magnetic field) \cite{Komargodski2017} and recent follow-up work \cite{Argurio2019,Baumgartner2019,Armoni2019,Baumgartner2020}, in which it has been suggested that the Yang-Mills term can play a non-trivial role despite superficially appearing irrelevant (in the sense of the renormalization group).  %they relate. 

This work is organized as follows. In Section \ref{sec:duality-review}, we review the non-Abelian Chern-Simons-matter dualities, focusing on the quadrality in Eq. \eqref{eq: SU(2) dualities, intro}, and we review the construction of Ref. \cite{Goldman2019}. % $SU(2)$ string of dualities relevant to us.  
%We also describe the phase diagrams of its constituent bosonic theories. % that appear in this string. 
In Section \ref{sec:fermions}, we present a detailed description of the dual fermionic theories and the properties of their IQH states, and we present the cases in which they appear to predict different topological orders. %when a magnetic field is turned on, emphasizing the paradoxical situations in which the two theories predict different topological orders. In the subsequent section, 
We then present our proposed scenario for how this state of affairs can be made consistent and provide additional examples of dual fermionic theories displaying similar phenomena. %\textcolor{violet}{In the process, we arrive at a new description an entire series of non-Abelian FQH states, among which is the of the anti-Pfaffian state \cite{Lee2007,Levin2007} as an IQH state of a non-Abelian composite fermion theory.} 
In the process, we uncover an entire series of non-Abelian FQH states, including the anti-Pfaffian \cite{Lee2007,Levin2007}, which we find can be simultaneously described as arising from IQH states of composite fermions in a non-Abelian theory and from pairing instabilities of a composite Fermi liquid in a dual Abelian theory. In Section \ref{sec:pairing}, %we address the original (but now secondary) aim of this work, which was to 
%having closed off the most immediate path to non-Abelian topological order more exotic than of the Moore-Read states, 
we demonstrate how the $U(k)_2$ states can be obtained through stacking and excitonic pairing in Abelian FQH states.  %in the dual fermionic language. 
Finally, we conclude with a discussion of our results and their implications.

\section{Review of non-Abelian dualities and the Landau-Ginzburg approach}
\label{sec:duality-review}
%To begin, 

\subsection{Non-Abelian dualities and the $\nu=1/2$ bosonic Laughlin state}
In this section, we first briefly review the dualities relevant to describing the $\nu=1/2$ bosonic Laughlin state. The recently proposed non-Abelian Chern-Simons-matter dualities relate theories of Wilson-Fisher bosons coupled to a Chern-Simons gauge field to theories of Dirac fermions also coupled to a Chern-Simons gauge field, with the matter content in the fundamental representation of the gauge group. We can write these dualities schematically as
\bea
\label{eq: U/SU}
\text{$N_f$ scalars + $U(N)_{k,k}$} &\longleftrightarrow& \text{$N_f$ fermions + $SU(k)_{-N+N_f/2}$}\,, \\
\label{eq: SU/U}
\text{$N_f$ scalars + $SU(N)_k$} &\longleftrightarrow& \text{$N_f$ fermions + $U(k)_{-N+N_f/2,-N+N_f/2}$}\,,\\
\label{eq: U/U}
\text{$N_f$ scalars + $U(N)_{k,k+N}$} &\longleftrightarrow& \text{$N_f$ fermions + $U(k)_{-N+N_f/2,-N-k+N_f/2}$}\,.
\eea
Our conventions and notation concerning non-Abelian Chern-Simons theories are presented in  Appendix \ref{sec:appendix-conventions}. 

%Of interest to us is the following subset of dualities,
The dualities, Eqs. \eqref{eq: U/SU}-\eqref{eq: U/U}, can be shown to imply the quadrality described in the Introduction \cite{Hsin2016},
\begin{align}
\text{a scalar}+U(1)_{2}&\longleftrightarrow\text{a fermion}+SU(2)_{-1/2}\nonumber\\
\label{eq: SU(2) dualities}
\updownarrow\hspace{1.2cm}&\\
\text{a fermion}+U(1)_{-3/2}&\longleftrightarrow\text{a scalar}+SU(2)_1\,.\nonumber
\end{align}
%where \textbf{Theory A} 
The first theory (top left) is the relativistic version of the usual Landau-Ginzburg theory for the $\nu=1/2$ bosonic Laughlin state. Explicitly, it is described by the Lagrangian,
\be
\label{eq: U(1)2}
\mathcal{L}_\Phi=|D_a\Phi|^2-|\Phi|^4+\frac{2}{4\pi}ada+\frac{1}{2\pi}Ada \, .
\ee
Here $a$ is an emergent $U(1)$ gauge field; $A$ the background electromagnetic (EM) field; we use the notation $D_a^\mu=\partial^\mu-ia^\mu$ and $adb=\varepsilon^{\mu\nu\lambda}a_\mu\partial_\nu b_\lambda$; and the term $-|\Phi|^4$ %is shorthand to 
denotes tuning to the Wilson-Fisher fixed point. The dual non-Abelian %\textbf{Theory B} 
bosonic theory (bottom right) is given by
\be
\label{eq: U(1)2 duality}
\mathcal{L}_\phi=|D_{u-A\mathbf{1}/2}\phi|^2-|\phi|^4+\frac{1}{4\pi}\Tr\left[udu-\frac{2i}{3}u^3\right]-\frac{1}{2}\frac{1}{4\pi}AdA\, ,
\ee
where $u$ is an $SU(2)$ gauge field, $\mathbf{1}$ is the $2\times 2$ identity matrix in color space, and $-|\phi|^4$ again denotes tuning to the Wilson-Fisher fixed point. By turning on mass operators, both theories can be shown to describe the transition between the $\nu=1/2$ bosonic Laughlin state and the trivial insulator, which correspond to their gapped ($\langle \Phi \rangle = 0$, $\langle \phi \rangle = 0$) and condensed ($\langle \Phi \rangle \neq 0$, $\langle \phi \rangle \neq 0$) phases, respectively. Essential to this conclusion is the fact that $SU(2)_{1}$, the TQFT obtained when $\Phi$ is gapped, is Abelian at the level of the braid group: it is equivalent to $U(1)_2$ by level-rank duality. 

The $SU(2)$ theory, $\mathcal{L}_\phi$, served as the main building block in our construction of the Read-Rezayi states in Ref. \cite{Goldman2019}, in which we %``stacked" 
considered multiple layers of the $\nu=1/2$ bosonic Laughlin state and introduced an interlayer pairing interaction for the $\Phi$ particles. %in the non-Abelian composite boson language of \textbf{Theory B}. 
The paired phase of this theory yielded the Read-Rezayi states. However, in that work we did not consider the landscape of non-Abelian phases accessible by dual theories of Dirac fermions, which we now turn to. %We next discuss the dual fermionic theories, of which we will make later use in a similar construction of the (time reversed) Read-Rezayi states.

\subsection{A comment on level-rank duality and topological orders of fermions}
Before describing the composite fermion theories of interest, we mention here a subtlety that arises when considering topological orders of composite fermions. When assessing the anyon content of the corresponding gauge theory, it is necessary to account for the the fact that the degrees of freedom charged under the gauge field are fermions, which affects the statistics of certain anyons by a minus sign. More technically, this is a result of the fact that gauge fields which couple to fermions are spin (actually spin$_c$, see Appendix \ref{sec:appendix-conventions}) connections, as opposed to the $U(1)$ connections that couple to bosons. We will refer to such gauge fields throughout this paper as spin gauge fields, and we will denote their associated TQFTs with the superscript `spin.' In general, level-rank duality can be thought of as relating a TQFT with a spin gauge field (spin TQFT) to one with a $U(1)$ gauge field.\footnote{Level-rank duality can be equivalently formulated to relate two spin-TQFTs by adding an invisible spin-1/2 line (also known in the condensed matter literature as a local spin-1/2 particle) to each side of the duality \cite{Hsin2016}. This formulation is less physical if we wish to view the fundamental charges at short distances as bosons, so we will refrain from using it.} Therefore, since composite fermion theories give rise to spin TQFTs, we will frequently invoke level-rank duality below and refer to a state's topological order via its corresponding (non-spin) TQFT. For example, if a composite fermion theory yields a $SU(2)_{-k}^{\mathrm{spin}}$ TQFT, we will refer to the associated topological phase by its level-rank dual, $U(k)_2$.

This formal discussion has physical implications. For example, consider the topological order of the $\nu=1/2$ Laughlin state. The anyons of this state are semions, with $\pi/2$ statistics. This state can be equally well described by a $U(1)_2$ TQFT or a $U(1)_{-2}$ TQFT with a spin gauge field, which we will denote $U(1)_{-2}^{\mathrm{spin}}$. As we will see below, this theory arises on integrating out a Landau level of composite fermions. While it appears that the anyons in this theory are antisemions (statistics $-\pi/2$), the $\pi$ statistics of the composite fermions converts them into semions. This is a type of level-rank duality, relating $U(1)_2$ to $U(1)_{-2}^{\mathrm{spin}}$. In this sense, level-rank dualities can generally be viewed as boson-fermion dualities, with some interesting exceptions, e.g. in the $SU(2)_1\leftrightarrow U(1)_2$ duality mentioned above, neither theory is spin.

\section{Non-Abelian dualities and the dynamics of composite fermions}
\label{sec:fermions}
%\subsection{An Abelian/non-Abelian fermionic duality}% and their integer quantum Hall phases}
%We now turn to the focus of this work, the fermionic theories. \textbf{Theory A$'$} is given by 
\subsection{The $\nu=1/2$ Laughlin state and a non-Abelian fermion-fermion duality} 

The bosonic theories described above are dual to a theory of Dirac fermions coupled to a $U(1)_{-3/2}$ Chern-Simons gauge field. For clarity, we will refer to this theory as \textbf{Theory A},
\be
\mathcal{L}_A= i\bar\psi\slashed{D}_a\psi-\frac{3}{2}\frac{1}{4\pi}ada - \frac{1}{2\pi}adA - \frac{1}{4\pi} AdA+\cdots \, , \label{eqn:TheoryC}
\ee
where $a$ is a $U(1)$ gauge field and we use the notation $\slashed{D}=D_\mu\gamma^\mu$, where $\gamma^\mu$ are the Dirac gamma matrices. This theory is also  dual to a theory of Dirac fermions coupled to a non-Abelian, $SU(2)_{-1/2}$ gauge field, which we will refer to as \textbf{Theory B},
\be
\mathcal{L}_B= i\bar\chi\slashed{D}_{b- A \mathbf{1}/2}\chi - \frac{1}{2} \frac{1}{4\pi} \mathrm{Tr}\left[ bdb - \frac{2i}{3} b^3 \right] - \frac{1}{4} \frac{1}{4\pi} AdA+\cdots \, , \label{eqn:TheoryD}
\ee
where $b$ is an $SU(2)$ gauge field and $\mathbf{1}$ is the $2\times 2$ identity matrix.\footnote{Throughout this work, we approximate the Atiyah-Patodi-Singer $\eta$-invariant as a level-$\frac{1}{2}$ Chern-Simons term and explicitly include it in the Lagrangian.} %({\hart We should probably rename the gauge fields so that $a$ is the gauge field in Theory $a$ and $b$ is the gauge field in Theory B. -- HG}) \textcolor{violet}{(RS: done)} . 
Here the $\chi$ fields transform as a doublet under $SU(2)$, and they have charge $-1/2$ under the global EM symmetry, $U(1)_{\mathrm{EM}}$. The fundamental (unit) charges are therefore the baryons, $\varepsilon_{\alpha\beta}\chi^{\alpha}\chi^\beta$, where $\alpha,\beta=1,2$ are $SU(2)$ color indices. %\footnote{Note also that, in writing down Eq. \eqref{eqn:TheoryD}, we have suppressed the contribution of some additional $U(1)$ gauge fields that do not couple to the rest of the theory. These provide compensating spin-1/2 line operators that are required for the theories in Eq. \eqref{eq: SU(2) dualities} to host matching topological orders \cite{Hsin2016}.}.   %We will refer to the $\psi$ and $\chi$ fields \sout{Abelian and non-Abelian} \textcolor{blue}{as the  composite fermions of the $U(1)$ and $SU(2)$ gauge theories}, respectively. 
%Note that 
Finally, the ellipses refer to irrelevant operators, such as Maxwell or Yang-Mills terms for the gauge fields. These operators are normally dropped since the duality is only valid in the IR limit, in which these operators are taken to zero, and their usual purpose is to provide UV regularization. However, we will see in the sections below that these operators can play important roles in determining low energy physics when background fields are turned on.

Being dual to the bosonic theories discussed in Section 2, \textbf{Theory A} and \textbf{Theory B} each describe a transition from the $\nu=1/2$ bosonic Laughlin state, which has $U(1)_2$ topological order, to a trivial insulator. This can be seen by introducing mass terms, $-m_\psi\bar\psi\psi$ and $-m_\chi\bar\chi\chi$, to their respective theories. For $m_\psi>0,m_\chi>0$, integrating out $\psi$ and $\chi$ can be seen to immediately yield an insulating state with vanishing Hall conductivity, $\sigma_{xy}=0$. On the other hand, when $m_\psi<0,m_\chi<0$, integrating out the composite fermions yields a state with $\sigma_{xy}=-\frac{1}{2}\frac{1}{2\pi}$ ($m_\psi<0,m_\chi<0$), with \textbf{Theory A} yielding a $U(1)_{-2}^{\mathrm{spin}}$ gauge theory and \textbf{Theory B} yielding $SU(2)_{-1}^{\mathrm{spin}}$. These constitute the same topological order as the $U(1)_2$ state by level-rank duality.  

By differentiating this pair of Lagriangians with respect to the background EM gauge field, $A_\mu$, to obtain the global EM charge current, $J^\mu$, one observes that, under the duality, the monopole current of \textbf{Theory A} is related to the baryon number current of \textbf{Theory B},
\be
\label{eq: current dictionary}
J_e^\mu=\frac{1}{2\pi}\,\varepsilon^{\mu\nu\lambda}\pd_\nu (a_\lambda -A_\lambda) \leftrightarrow  -\frac{1}{2}\, j^\mu_\chi-\frac{1}{2}\frac{1}{2\pi}\varepsilon^{\mu\nu\lambda}\partial_\nu A_\lambda\,,
\ee
where $j^\mu_\chi=\bar\chi\gamma^\mu\chi$ is the $\chi$ charge current of the \textbf{Theory B}. The interpretation of this dictionary is analogous to charge-vortex duality \cite{Peskin1978,Dasgupta1981}: flux of the gauge field $a$ in \textbf{Theory A} maps to charge of the $\chi$ fermions in \textbf{Theory B}. The same interpretation applies to the pair of bosonic theories discussed in the previous Section.
%where $\gamma^\mu$ are the gamma matrices. \textcolor{violet}{(Maybe this equation is not needed)}

\subsection{Abelian and non-Abelian Jain sequences}
\label{sec: Jains}
%In contrast to their bosonic duals, we can more readily access a wider range of FQH states in \textbf{Theories C} and \textbf{D}, beyond simply the $\nu=0$ and $\nu=1/2$ states, 
In contrast to their bosonic counterparts, these  composite fermions each satisfy the Pauli exclusion principle. As a result, it is natural to consider the gapped phases accessible by filling up Landau levels and forming IQH states, in analogy to the construction of the Jain sequences, in which FQH phases are obtained as IQH states of composite fermions \cite{Jain-1989}. % of the composite fermions. 
In particular, integer quantum Hall states of the $SU(2)$ doublet composite fermions, $\chi$, can be expected to yield non-Abelian topological orders. This method of forming non-Abelian quantum Hall phases appears to be quite natural, but we will quickly learn that the non-Abelian dualities imply that things are not so simple, and the ultimate choice of ground state will be sensitive to the order in which the lowest Landau level and IR limits are taken. %more immediate than the multilayer pairing constructions applied to the bosonic theories. %However, we will quickly find that \textcolor{violet}{taking the former route will lead us to interesting conclusions about the dynamics of the Abelian and non-Abelian fermionic theories related by duality.} \sout{a richer set of outcomes is possible.}

To this end, we begin by relating the electronic filling fraction, $\nu$, to the filling fractions of the $\psi$ and $\chi$ fermions using the dictionary, Eq. \eqref{eq: current dictionary}. Focusing first on the Abelian \textbf{Theory A}, the physical electric charge density is given in terms of the magnetic flux felt by the composite fermions,  %we compute the electronic charge density, $\rho_e$, to be
\begin{align}
\label{eq: rhoe Theory A}
\rho_e = \left\langle J^0_e\right\rangle = - \frac{1}{2\pi} \langle\varepsilon^{ij}\pd_i a_j\rangle - \frac{1}{2\pi} B,
\end{align}
where $B=\varepsilon^{ij}\partial_i A_j$ is the background magnetic field. We use brackets here to emphasize that that we define $\rho_e$ to be the expectation value of the charge density operator. %The equation of motion for $b_t$ yields 
We can relate $\rho_e$ to the composite fermion charge density through the equation of motion for $a_0$,
\begin{align}
\label{eq: bt EoM Theory A}
0  = \langle\psi^\dagger\psi\rangle - \frac{3}{4\pi}\langle\varepsilon^{ij}\pd_i a_j\rangle - \frac{1}{2\pi}B \, ,
\end{align}
%where $\rho_\psi=\langle\psi^\dagger\psi\rangle$ is the $\psi$  density. 
%Defining $\rho_\psi=\langle\psi^\dagger\psi\rangle$ to be the composite fermion density of \textbf{Theory A}, we see that the presence of a non-zero flux of $b_\mu$, $b_*=\langle \varepsilon^{ij}\pd_ib_j\rangle$, will yield Landau levels of the $\psi$-fermions, with filling fraction
If we define the composite fermion filling fraction of $\textbf{Theory A}$ to be
\begin{align}
\nu_\psi = 2\pi\,\frac{\langle\psi^\dagger\psi\rangle}{\langle\varepsilon^{ij}\partial_i a_j\rangle},
\end{align}
%Using this definition and the above equations, we find the following relation between the electronic and $\psi$-fermion filling fractions:
we obtain a relation between the composite fermion and electronic filling fractions,
\begin{align}
\nu =-2\pi\,\frac{\rho_e}{B}=  \frac{\nu_\psi - 1/2}{\nu_\psi - 3/2}\,. \label{eqn:nu_e-nu_psi}
\end{align}
Note that we have absorbed a minus sign into the definition of $\nu$ for notational convenience. From this formula, we see that IQH states of the $\psi$ fermions, which occur at fillings $\nu_\psi=p-1/2$, correspond to the known (descendent) bosonic Jain sequence states,
\be
\label{eq: Abelian Jain}
\nu_p=\frac{p-1}{p-2}\,,\,p\in\mathbb{Z}\,.
\ee 
Indeed, integrating out the composite fermions yields the Lagrangian,
\be
\mathcal{L}_{A,\mathrm{eff}} = \frac{p-2}{4\pi}ada - \frac{1}{2\pi}adA - \frac{1}{4\pi} AdA. \label{eqn:TheoryC-eff}
\ee
Each of these states (with the exception of the states at $p=1,2$, which are respectively a trivial insulator and a superfluid) is an Abelian FQH phase of the physical charges, which here are bosons.

%Turning now to \textbf{Theory B$'$}, 
The same type of analysis can be carried out for \textbf{Theory B}, leading to a non-Abelian version of the bosonic Jain sequence. %We see that the $\chi$-fermions couple directly to the EM field. Hence 
Recalling Eq. \eqref{eq: current dictionary}, the electric charge density is directly related to the density of $\chi$ fermions, $\chi^\dagger\chi$, via
\begin{align}
\label{eq: rhoe to rho chi}
\rho_e = - \frac{1}{2} \langle\chi^\dagger\chi\rangle - \frac{1}{4} \frac{1}{2\pi} B.
\end{align}
%We assume that the net non-Abelian flux vanishes, $\langle (da)_t \rangle = 0$ and that 
%Here we make the crucial assumption that the $\chi$'s are deconfined, since they are coupled to a Chern-Simons gauge field.
Because the $\chi$ fermions are coupled to a Chern-Simons gauge field, they do not confine, meaning that a non-zero magnetic field, $B$, will %therefore result in the $\chi$-fermions forming 
cause them to form Landau levels with degeneracy, %In order to define the $\chi$ filling fraction, we first identify the Landau level degeneracy as being
\begin{align}
d_{LL} = \frac{BA}{2\pi}\times |q_\chi| \times (\text{color degeneracy}) = \frac{BA}{2\pi} \, , %\label{eqn:nu_chi}
\end{align}
where $A$ in this expression is the area of the system and $q_\chi=-1/2$ is the EM charge of the $\chi$ fermions. Therefore, the filling fraction of the $\chi$ fermions is 
\begin{align}
\nu_\chi = 2\pi\,\frac{\langle\chi^\dagger\chi\rangle}{B}\,. \label{eqn:nu_chi}
\end{align}
Plugging this into Eq. \eqref{eq: rhoe to rho chi} yields a relation between $\nu$ and $\nu_\chi$,
\begin{align}
\nu = \frac{1}{2} \nu_\chi + \frac{1}{4}. \label{eqn:nu_e-nu_chi}
\end{align}
When the $\chi$ fermions fill an integer number of Landau levels, $\nu_\chi=s-1/2$, and the filling of the physical charges is
\be
\label{eq: non-Abelian Jain}
\nu_s=\frac{s}{2}\,,s\in\mathbb{Z}\,.
\ee
On integrating out the composite fermions, one obtains a $SU(2)^{\mathrm{spin}}_{-s}$ theory with Lagrangian, %\textcolor{violet}{(RS: I think we need to be careful about the topological order when looking at Chern-Simons theories coupled to fermionic matter. The low energy theory is $SU(2)_{-s}$ but, loosely speaking, the probes are fermionic Wilson lines, so we need to supplement the spin of each anyon in $SU(2)_{-s}$ with an appropriate fermionic minus sign. Equivalently, one can look at the low energy theory as an $SU(2)_{-s}$ Chern-Simons theory of a spin gauge field. This is the perspective taken in Ref. \cite{Ma2020}. To identify the correct topological order, one can employ a level-rank duality to re-express the theory in terms of a usual gauge field. For instance, $SU(2)_{-s} \leftrightarrow U(s)_2$. The same issue arises in the theories of the $SU(2)$ quadrality. The gapped phases of the $\Phi$ and $\phi$ theories have $U(1)_2$ and $SU(2)_1$ topological order, which are the same and consist of the vacuum and the semion. The fermionic duals appear to predict $U(1)_{-2}$ and $SU(2)_{-1}$, which contain the vacuum and anti-semion; one needs to take into account that these are spin gauge fields / coupled to fermions / the presence of invisible lines to match up the topological order with that predicted by the bosonic theories.)} \textcolor{blue}{EF: we need to be careful with the terminology; I assume that what is being referred here as a ``spin gauge field" is a spin$_c$ connection.}   
\begin{align}
\mathcal{L}_{B,\,\mathrm{eff}} %&= - \frac{s-1/2}{4\pi} \Tr\left[(a - \frac{A\mathbf{1}}{2})d(a -\frac{A\mathbf{1}}{2}) - \frac{2i}{3}(a -\frac{A\mathbf{1}}{2})^3 \right] - \frac{1}{2} \frac{1}{4\pi} \mathrm{Tr}\left[ ada - \frac{2i}{3} a^3 \right] - \frac{1}{4} \frac{1}{4\pi} AdA \\
&= - \frac{s}{4\pi} \mathrm{Tr}\left[ bdb - \frac{2i}{3} b^3 \right] - \frac{s}{2} \frac{1}{4\pi} A d A\,,
\end{align}
which describes a non-Abelian, $U(s)_2$, topological order when $|s|>1$. This approach recalls earlier approaches to non-Abelian FQH states using parton constructions \cite{Wen1991,Wen1999,Barkeshli2010}. However, unlike in those cases, in which the elctron operator is fractionalized by hand and it is necessary to require that the partons do not confine by fiat, here the duality provides a clear basis for the presence of a (deconfining) non-Abelian gauge field.

%Now, if at a given electronic filling, $\nu$, {\it both} the \sout{Abelian} \textcolor{blue}{$\psi$} composite fermions \sout{, $\psi$,} and the \sout{non-Abelian} \textcolor{blue}{$\chi$}  composite fermions \sout{, $\chi$,} form IQH states, there may be cause for concern since it would appear that the latter state could possess non-Abelian topological order. Indeed, if the $\chi$ particles fill $s$ LLs, we can integrate them out, yielding a theory which would seem to have $SU(2)_{-s}$ topological order. To determine whether this situation arises here, we combine Eqs. (\ref{eqn:nu_e-nu_psi}) and (\ref{eqn:nu_e-nu_chi}) to relate the $\chi$ and $\psi$ filling fractions:
%\begin{align}
%\nu_\chi = \frac{3\nu_\psi - 1/2}{2\nu_\psi - 3}. \label{eqn:nu_chi-nu_psi-relation}
%\end{align}
%We now assume that both \sout{the Abelian and non-Abelian} \textcolor{blue}{$\psi$ and $\chi$ composite fermions} fill up an integer number of Landau levels, so that $\nu_\psi = p - 1/2$ and  $\nu_\chi = s-1/2$, with $p,s \in \mathbb{Z}$. Plugging these expressions into Eq. (\ref{eqn:nu_chi-nu_psi-relation}), we obtain the relation

Having obtained the sequences of incompressible filling fractions associated with \textbf{Theory A} and \textbf{Theory B}, one can immediately observe several special (non-trivial) filling fractions, $\nu_*$, at which \emph{they coincide}, indicating the presence of competing ground states. Comparing Eqs. \eqref{eq: Abelian Jain} and \eqref{eq: non-Abelian Jain}, these fillings occur when $\nu_p=\nu_s=\nu_*$, i.e.
\begin{align}
s = 2 \left(1 + \frac{1}{p-2} \right)\,,\,p,s\in\mathbb{Z}. \label{eqn:s-p-relation}
\end{align}
%Requiring that both $p$ and $s$ be integer and finite implies that 
Here we recall that $p$ and $s$ are the number of Landau levels filled by the $\psi$ and $\chi$ fermions, respectively. This equation has several solutions, %$p-2 = \pm 1, \pm 2$. 
which are organized in Table \ref{tab:shared-IQH-states}. %lists these four solutions, along with the corresponding values of $s$ and the electronic filling fraction $\nu$. 
Two of these solutions, $(p=1,s=0)$ and $(p=0,s=1)$, respectively correspond to the $\nu=1/2$ bosonic Laughlin state, which has $U(1)_{2}\leftrightarrow SU(2)_{-1}^{\mathrm{spin}}$ topological order. This is consistent with the fact that at criticality these theories describe a plateau transition between these states.

It is the presence of other solutions, at $(p=3,s=4)$  and $(p=4,s=3)$, corresponding to $\nu_*=2$ and $\nu_*=3/2$ respectively, that reveals new physics. %the interpretation of the $p=3,4$ states, corresponding to $\nu=-2,-3/2$, is not so clear. 
On the one hand, \textbf{Theory A} predicts the $\nu_*=3/2$ and $\nu_*=2$ states to have $U(1)_{-2}$ and $U(1)_1$ (i.e. trivial) topological orders. On the other hand,  \textbf{Theory B} predicts the \emph{same} states to have \emph{non-Abelian} $SU(2)_{-3}^{\mathrm{spin}}\leftrightarrow U(3)_2$ and $SU(2)_{-4}^{\mathrm{spin}}\leftrightarrow U(4)_2$ topological orders. %Demonstrating the existence of this apparently severe inconsistency is one of the main messages of the present work. The next order of business is to provide a physically reasonable resolution of this discrepancy.%, to which we turn next.
While it is common in quantum Hall physics for different competing states to be proposed for the same filling fraction, the duality of \textbf{Theory A} and \textbf{Theory B} identifies the two theories' ground states. Therefore, the consistency of the duality implies that the conditions under which the $\psi$ fermions form an IQH state are not the same as those of the $\chi$ fermions, and the two possible states must be separated by a phase transition. A theory of this phase transition requires short-distance dynamical information not specified by, but consistent with, the duality. We now present a possible scenario for a transition of this kind.
%\sout{We now proceed to describe one} \textcolor{blue}{In the next subsection we present a possible scenario for a transition of this kind.} \sout{in the next subsection, in which the aforementioned ellipses in Eqs. \eqref{eqn:TheoryC} and \eqref{eqn:TheoryD} play an essential role. We emphasize that dynamical information of this kind -- which we will see below involves strongly coupled physics -- is unprecedented in the non-perturbative study of dualities in quantum field theory, and it represents new information about the dynamics of Chern-Simons-matter theories in the presence of background fields} ({\hart something like this paragraph should tie up the introduction --HG}). 

\begin{table}
 \centering
 \large
 \begin{tabular}{||c || c | c | c | c|} 
 \hline
  $\nu_*$ & $0$ & $\frac{1}{2}$ & $\frac{3}{2}$ & $2$\\
 \hline
 $\nu_\psi+1/2$ & $1$ & $0$ & $4$ & $3$ \\ %
 \hline
 $\nu_\chi+1/2$ & $0$ & $1$ & $3$ & $4$ \\ 
 \hline
 \textbf{Theory A} & Trivial & $U(1)_{2}$ & $U(1)_{-2}$ & IQH \\
 \hline
 \textbf{Theory B} & Trivial & $U(1)_2$ & $U(3)_2$ & $U(4)_2$ \\
 \hline
\end{tabular} 
 \caption{Solutions to Eq. (\ref{eqn:s-p-relation}), in which both the $\psi$ and $\chi$ composite fermions form IQH states at special electronic filling fractions, $\nu_*$. Also indicated are the topological orders predicted by the dual theories for each filling. %({\hart I have re-named the state predicted by Theory A for $\nu_*=2$ an IQH state since it has integer Hall conductivity. It should be possible to check whether this state has the right gravitational anomaly. -- HG}). %electronic filling fractions $\nu$ and the topological orders predicted by \textbf{Theories C} and \textbf{D} at these fillings.}
 }
 \label{tab:shared-IQH-states}
\end{table}

\subsection{Dynamical scenario}
\label{sec: dynamical scenario}

We now provide a possible explanation of the physics occurring at the special filling fractions $\nu_*=3/2,2$. For now, we will work from the point of view of the non-Abelian \textbf{Theory B}, and we will begin by considering what happens as we fill Landau Levels. %scenario in which the predictions of \textbf{Theory A} and \textbf{Theory B} for the nature of the states at electronic filling fractions $\nu_*=3/2,-2$. %can be made consistent. 
From Table \ref{tab:shared-IQH-states}, we see %that when the %two theories predict the same states when the %non-Abelian composite 
that filling the zeroth and first Landau Levels of the $\chi$ fermions %of \textbf{Theory B} fill zeroth or one Landau level. 
corresponds to the expected trivial insulator ($\nu=0$) and bosonic Laughlin $(\nu=1/2)$ states. What occurs when the non-Abelian composite fermions fill two Landau levels is also quite non-trivial, but we will table that discussion until the next subsection.  %to this question later in this section. %When the non-Abelian composite fermions fill two Landau levels, \textbf{Theory B$'$} predicts a state with $SU(2)_{-2}$ topological order, while \textbf{Theory A$'$} predicts a gapless state, which may plausibly have an instability to a paired state possessing the same topological order. 
%However, the primary trouble 
For now, our concern will be what happens when we fill the third Landau level, corresponding to $\nu_*=3/2$. Our proposal for $\nu_*=2$ will prove to be essentially identical. At $\nu_*=3/2$, \textbf{Theory B} predicts an incompressible state with $SU(2)_{-3}^{\mathrm{spin}} \leftrightarrow U(3)_2$ topological order, while \textbf{Theory A} predicts $U(1)_2$. %of the \sout{non-Abelian} composite fermions \textcolor{blue}{of the non-Abelian theory}. 
This suggests that it should be possible to trigger an instability in the non-Abelian theory as we fill this Landau level, which preempts the $U(3)_{2}$ topological order and yields the same Abelian phase predicted by \textbf{Theory A} (and vice versa). %In other words, our claim is that 
Consequently, both the $U(1)_{2}$ and $U(3)_{2}$ phases must exist in the $\nu=3/2$ phase diagram. This is the only state of affairs consistent with the duality. %We now argue how these two phases should appear in the phase diagram in the language of \textbf{Theory B}.

\begin{comment}
\begin{figure}
\centering
\includegraphics[width=0.65\textwidth]{SU(2)3_phase_diagram.png}
\caption{Proposed phase diagram for the dimensionless coupling, $\bar\lambda=g_{YM}^2/\omega_c$. When $\bar\lambda\rightarrow\infty$, the Yang-Mills term vanishes, and the picture of deconfined $\chi$ fermions filling Landau levels is valid, leading to the $SU(2)_{-3}$ state. When $\bar\lambda$ runs small, the Yang-Mills term becomes very large, Landau levels can mix, and the deconfinement of $\chi$ charges is no longer assured, leading ultimately to the $U(1)_2$ phase predicted in \textbf{Theory A}. These states are separated by a critical point at $\bar\lambda=\bar\lambda_*\sim\mathcal{O}(1)$, which is likely first order.} 
\label{fig: SU(2)3 phase diagram}
\end{figure}
\end{comment}

How might such an instability occur? It is here that the ellipses in Eqs. \eqref{eqn:TheoryC} and \eqref{eqn:TheoryD} become crucial.\footnote{We thank Chong Wang for enlightening discussions on this point.} These ellipses include operators that are irrelevant at tree level but may nevertheless play an important role in determining the low energy physics when a magnetic field and chemical potential are introduced. Indeed, there is no sense in which such fields are ever perturbative, as they reorganize the spectrum of a theory in dramatic ways. To make this discussion more precise, consider the Yang-Mills term in \textbf{Theory B},
\be
\mathcal{L}_{YM}=-\frac{1}{2g_{YM}^2}\Tr[f_{\mu\nu}f^{\mu\nu}]\,,
\ee
where $f_{\mu\nu}=\partial_\mu b_\nu-\partial_\nu b_\mu-i[b_\mu,b_\nu]$ is the field strength of the $SU(2)$ gauge field, $b$. At tree level, the mass dimension of the operator $\Tr[f^2]$ is 4, meaning that $[g_{YM}^2]=1$: it is an energy scale. Commonly, the IR limit in which the duality of Eqs. \eqref{eqn:TheoryC} and \eqref{eqn:TheoryD} holds is phrased as the limit $g_{YM}^2\sim\Lambda\rightarrow\infty$, where $\Lambda$ is a UV cutoff, but strictly speaking this is only true in the absence of a background magnetic field, which provides its own energy scale in the form of the cyclotron frequency, $\omega_c\sim \sqrt{B}$ (for massless Dirac fermions). 

Consequently, one can form a dimensionless coupling,
\be
\label{eq: lambda-bar}
\bar\lambda=\frac{g_{YM}^2}{\omega_c}\,,
\ee
that can have non-trivial running as a result of strong interaction effects. This would mean that the ground state ultimately chosen by \textbf{Theory A} can depend on the order of the limits, $g_{YM}^2\rightarrow\infty$ and $\omega_c\rightarrow\infty$. Importantly, one order of limits in \textbf{Theory B} may not correspond to an analogous order of limits in \textbf{Theory A}. In other words, \textbf{Theory B} may be in a strongly coupled regime ($0\leq\bar\lambda<\infty$) while $\textbf{Theory A}$ may behave as if all of the irrelevant operators have been taken to zero prior to taking $\omega_c\rightarrow\infty$. 

%Indeed, there are two energy scales present in \textbf{Theory B$'$} in the presence of a non-zero magnetic field: the cyclotron frequency, $\omega_c$, and the Yang-Mills coupling, $g_{YM}^{2}$ (recall that a Yang-Mills term is always implicit in the action -- see Appendix \ref{sec:appendix-conventions}). From these scales, we can form the dimensionless ratio $\lambda = g_{YM}^2/\omega_c$. As the only such parameter in our theory, for our picture to hold, a transition between the $U(1)_2$ and $SU(2)_{-3}$ topological phases must occur as $\lambda$ is varied. 

This is the essence of our proposal\footnote{We emphasize that while we focus on the example of the Yang-Mills term, there are many other operators that might be responsible for the behavior we propose, and it may be more correct to consider a linear combination of these operators as being responsible. Another likely family of examples is four-fermion operators, which have the same tree level dimension as the Yang-Mills operator.} % Such operators are known to have interesting scaling behavior in QED$_3$ and QCD$_3$ {(\hart cite self, Komargodski et al., Pufu et al.}), albeit in the absence of background fields and Chern-Simons terms.} 
for the phase diagram of $\textbf{Theory B}$, shown schematically in Fig. \ref{fig: SU(2)3 phase diagram}. Whether the theory is in the $U(3)_{2}$ or $U(1)_2$ phase is determined by the value of $\bar\lambda$, with the two phases being separated by a phase transition at a value $\bar\lambda=\bar\lambda_*\sim\mathcal{O}(1)$. %To see where these phases should lie on the phase diagram, we first note that 
For $\bar\lambda > \bar\lambda_*$, $\bar\lambda$ runs large, corresponding to the limit $g_{YM}^2\rightarrow\infty$ followed by $\omega_c\rightarrow\infty$. In this phase, the Yang-Mills term disappears, leaving the Chern-Simons term, which ensures that the composite fermions, $\chi$, are deconfined. It is in this regime that we expect the picture described in the previous subsection of deconfined $\chi$ fermions filling  Landau levels to hold, making this the phase with $U(3)_{2}$ topological order. On the other hand, for $\bar\lambda < \bar\lambda_*$, we propose that $\bar\lambda$ runs small. Here the Yang-Mills term becomes important, and the assumption of deconfined composite fermions breaks down: the Landau levels mix. As a result, the composite fermions will tend to form   bound states that are neutral under $SU(2)$: the baryons, $\varepsilon_{\alpha\beta}\chi^\alpha\chi^\beta$. These are bosons of charge $-1$ since the fermions are doublets under $SU(2)$. Being the physical electric charges in this theory, these bosonic baryons will find themselves at filling $\nu=3/2$, and the resulting topological order will ultimately be $U(1)_{2}$, as predicted by \textbf{Theory A}. Note that this final conclusion is conjecture based on the consistency of the duality -- the physics in such a phase is very strongly coupled, and we cannot show explicitly that this is the true ground state. We also point out that $\bar\lambda$ may not run to zero in this phase, instead running to a small but finite value. This represents a very interesting but theoretically daunting possibility.

We now comment on what occurs in the dual description of $\textbf{Theory A}$, in which we can define a similar running coupling\footnote{Note that unlike the  fermions of \textbf{Theory B}, the magnetic field felt by the $\psi$ fermions depends both on the background chemical potential and magnetic field, so $\omega_c$ is not precisely the Landau level gap, which is determined by $\langle\varepsilon^{ij}\pd_i a_j\rangle$.}, $\bar\lambda_A=g_{\mathrm{Maxwell}}^2/\omega_c$, where $g_{\mathrm{Maxwell}}$ is the coupling associated with the Maxwell term for the $U(1)$ gauge field, $a$. In this theory, the limit $\bar\lambda_A\rightarrow\infty$ corresponds to the $U(1)_2$ (IQH) state at filling $\nu_*=3/2$ ($\nu_*=2$), as the Maxwell term vanishes. However, since the Maxwell term is not dual to the Yang-Mills term, it is not clear whether the phase transitions described for \textbf{Theory B} correspond to transitions tuned by $\bar\lambda_A$ or another coupling, such as one associated with a linear combination of four-fermion operators. Nevertheless, unless $\bar\lambda_*=\infty$ in \textbf{Theory B} (in which case the non-Abelian phase is nowhere stable, having no basin of attraction), the duality indicates that the non-Abelian  states should be accessible to \textbf{Theory A} as well, and that these transitions should also be present in that theory. %Note that unlike the  fermions of \textbf{Theory B}, the magnetic field felt by the $\psi$ fermions depends both on the background chemical potential and magnetic field, so $\omega_c$ is not precisely the Landau level gap, which is determined by $\langle\varepsilon^{ij}\pd_i b_j\rangle$.

%Finally, we consider $\nu_*=2$, in which the \textbf{Theory B} composite fermions fill four Landau levels, indicating $SU(2)_{-4}$ topological order. In this case, \textbf{Theory A} predicts an IQH (i.e. topologically trivial) state. Our proposal for this case proceeds in analogy with the above discussion of the states at $\nu_*=3/2$. %While a similar interpretation can also apply to this transition, we note that it is also possible to obtain this kind of transition from an off-diagonal condensate of fermion-antifermion pairs, e.g. $\langle\bar\chi_1\chi_2\rangle\neq 0$, where $1,2$ are color indices. This Higgses the $SU(2)$ gauge group while leaving $U(1)_{\mathrm{EM}}$ intact, resulting in the correct ground state. Again, it is not clear whether such a transition is continuous.

\subsection{Comments on the nature of the transition}

%What occurs between these two phases is by no means an easy question to address. \textcolor{blue}{The natural scenario is that both phases occur and that there is a phase transition at some critical value $\lambda_c$ of order 1. However, the nature of this phase transition depends on microscopic details.}
The nature of the transition between these phases  depends on microscopic details, and it is not immediately clear how to study the strongly coupled physics when $\bar\lambda\sim\mathcal{O}(1)$. One exciting possibility is that the phase transition is continuous, which would exist beyond the Landau paradigm since it separates two distinct topological orders. %\sout{there exists a continuous transition} \textcolor{blue}{the phase transition is continuous,}  \sout{between these two phases at some $\mathcal{O}(1)$ critical coupling, $\lambda = \lambda_c$}, 
Furthermore, this would imply the existence of an unstable conformal field theory (CFT) fixed point, which we expect would be quite exotic and perhaps involve emergent symmetries. %Such a critical theory would be quite exotic, describing the transition between two distinct topological orders. 
In the spirit of universality, we attempt here to write down the simplest possible theory of such a transition. The theory we find consists of $N_f=4$ flavors of electrically neutral Dirac fermions, $\xi_i$, coupled to a $SU(2)_{-1}^{\mathrm{spin}}$ Chern-Simons gauge field, $c$,
\be 
\mathcal{L}=\sum_{i=1}^{N_f=4}\bar\xi_i(i\slashed{D}_c-m_\xi)\xi_i-\frac{1}{4\pi}\Tr\left[cdc-\frac{2i}{3}c^3\right]-\frac{3}{2}\frac{1}{4\pi}AdA\,.
\ee
We emphasize that the fields $\xi$ and $c$ need not have any local relationship with the fields in \textbf{Theory A} nor \textbf{Theory B}. This theory has a $U(4)$ global symmetry rotating the fermion flavors. For $m_\xi \gg 0$, the theory is in the Abelian, $SU(2)_{+1}^{\mathrm{spin}}\leftrightarrow U(1)_{-2}$ phase, and for $m_\xi \ll 0$ the theory is in the non-Abelian, $SU(2)_{-3}^{\mathrm{spin}}\leftrightarrow U(3)_{2}$ phase. While the flavor index can be interpreted as a kind of Landau level index, there is no \emph{a priori} reason for this symmetry to be enforced, rendering this theory at best a multicritical point. What's more, this theory has $N_f>2|k|=2$, meaning that it falls outside of the Chern-Simons-matter dualities of Eqs. \eqref{eq: U/SU}-\eqref{eq: U/U} and may spontaneously break the flavor symmetry at small values of $m_\xi$ \cite{Komargodski2017}. We therefore find the presence of a direct second-order transition unlikely, and we conjecture that any other possible CFT with these two phases also has an enlarged set of global symmetries compared to the underlying UV problem.\footnote{While one may also wish to consider theories with bosonic matter, we note that it is not possible to condense bosonic operators to transition between $U(3)_2$ and $U(1)_{-2}$ due to the difference in the signs of the level. This cannot be repaired with level-rank duality because the $U(1)_{-2}$ gauge field is not spin.} We also conjecture that this is the case for the other transitions discussed in this work.

Therefore, it is perhaps more natural to expect the mundane scenario in which the two phases are separated by a first order transition. Starting in, say, the $U(1)_2$ phase, as $\bar\lambda$ is increased, phase separation will set in, yielding bubbles and stripes of the $U(3)_{2}$ phase, which eventually fill the system. %These two scenarios are illustrated schematically in Fig. (to be added {\hart (are we actually adding this figure? -- HG)}). 
It is also possible that the transition is not direct, and that several different phases arise when $\bar\lambda\sim\mathcal{O}(1)$. We cannot exclude this possibility. That we cannot present a thorough description of the transition is not surprising, given the complexity of the phase diagrams for electrons in magnetic fields at fractional filling. %However, we will present some speculative arguments in the next subsection based on effective field theory. %({\hart I have removed the triplet pairing example from this section because I'm no longer sure I believe it -- HG}).

\subsection{Non-Abelian duality and paired FQH phases}

\begin{table}
 \centering
 \large
 \begin{tabular}{||c || c | c |} 
 \hline
  $\nu_*$ & $-1$ & $\infty$\\
 \hline
 $\nu_\psi+1/2$ & $\infty$ & $2$  \\ %
 \hline
 $\nu_\chi+1/2$ & $2$ & $\infty$  \\ 
 \hline
 \textbf{Theory A} & Metal & Superfluid \\
 \hline
 \textbf{Theory B} & $U(2)_2$ & Metal  \\
 \hline
\end{tabular}
\caption{Solutions to Eq. \eqref{eqn:s-p-relation} in which one of the two dual theories is metallic, i.e. is at infinite filling.} 
\label{table: metallic fillings}
\end{table}

In the above subsection, we conspicuously left out a discussion of the case when the $\chi$ composite fermions fill two Landau levels, corresponding to filling $\nu_*=1$. At this filling, Eqs. \eqref{eq: rhoe Theory A} and \eqref{eq: bt EoM Theory A} indicate that the $\psi$ composite fermions of \textbf{Theory A} feel a vanishing magnetic field and therefore form a metallic state with density set by the background magnetic field, $\langle\psi^\dagger\psi\rangle=B/2\pi$. This is the well known metallic, composite Fermi liquid state of bosons with unit filling \cite{Read1998}. On the other hand, the analysis of Section \ref{sec: Jains} finds \textbf{Theory B} in an IQH state, yielding $U(2)_{2}$ topological order. Although one of the predicted phases is gapless, we nevertheless propose that the same scenario presented above holds in this theory: the ultimate choice of ground state is determined by the order in which the IR and $\omega_c\rightarrow\infty$ limits are taken. These states are again separated by a phase transition, which is tuned by the dimensionless coupling of the kind defined in Eq. \eqref{eq: lambda-bar}. 

In contrast to the cases in which both phases were gapped, here we can clearly understand this transition in terms of the $\psi$ composite fermions of \textbf{Theory A}. Indeed, the $U(2)_2$ state can be shown to be one of a range of non-Abelian phases of bosons at $\nu=1$ that can be obtained as paired states of the composite Fermi liquid \cite{Wang2016}, the most famous of which being the $SU(2)_2$ bosonic Pfaffian state \cite{Fradkin1998}. The major difference between the $U(2)_2$ state and the $SU(2)_2$ state lies in the topological spin of the non-Abelian half-vortex, which is set by the pairing momentum channel. As we will explain in more detail in the following subsection, the appropriate channel to obtain the $U(2)_2$ order is $l=2$. Consequently, this transition can be understood in terms of the flow of the pairing interaction, which is a four-fermion operator in the Abelian \textbf{Theory A}. Interestingly, it is expected from numerical simulations and recent analytic calculations that the composite Fermi liquid state of bosons at $\nu=1$ is unstable to pairing in the lowest Landau level limit \cite{Cooper2001,Dong2020}, perhaps suggesting that the metallic state seen in \textbf{Theory A} has no basin of attraction unless the theory is modified in some way.

From the point of view of \textbf{Theory B}, it is natural to expect that $\bar\lambda$ is again the correct running coupling, with the choice of ground state again being viewed as a question of the order in which the lowest Landau level and IR limits are taken. %for \textbf{Theory B}. %, this transition may be interpreted as follows. 
As above, for $\bar\lambda>\bar\lambda_*$, the $\chi$ fermions find themselves in an IQH state, yielding the $U(2)_{2}$ topological order. For $\bar\lambda < \bar\lambda_*$, this picture breaks down, leading to a theory involving bosonic baryons, which we conjecture form the $\nu=1$ metallic state.  

Remarkably, at filling $\nu_*=1$, we have observed a duality between composite fermion pairing in \textbf{Theory A} and the IQH effect in \textbf{Theory B}. This is surprising, as there is no known dictionary of local operators that makes this connection explicit. Indeed, while the $U(2)_2$ state has previously been obtined both in the Read-Green pairing picture \cite{Read2000,Fradkin1998} and as an IQH state of non-Abelian partons \cite{Wen1991,Wen1999,Barkeshli2010,Ma2020}, it has never before been suggested that these two constructions may be dual to one another. We will see below that this duality is not limited to the particular case of bosons at $\nu=1$: in Section \ref{sec:UN-duality-gapless-examples}, we will encounter a parallel story involving the anti-Pfaffian state of fermions at $\nu=1/2$.

Similar competing ground states are found as the external magnetic field is turned off, i.e. $\nu_*\rightarrow\infty$ (see Table \ref{table: metallic fillings}). %, with the roles of \textbf{Theory A} and \textbf{Theory B} being reversed (see Table \ref{table: metallic fillings}). 
In this case, the analysis of Section \ref{sec: Jains} indicates that \textbf{Theory A} predicts a superfluid state, the usual fate of bosons at finite density and $B=0$. On the other hand, the composite fermions of \textbf{Theory B} form a metal with density equal to the background charge density, a far more exotic non-Fermi liquid state which surely requires that the fundamental bosonic charges be very strongly interacting. Nevertheless, the interpretation of the transition between these states is natural from the point of view of \textbf{Theory B}: again adopting the notation of Section \ref{sec: dynamical scenario}, for $\bar\lambda > \bar\lambda_*$, the theory remains metallic, while for $\bar\lambda < \bar\lambda_*$, the $\chi$ fermions confine to form bosonic baryons. Since these bosons feel no magnetic field, they would then condense and spontaneously break $U(1)_{\mathrm{EM}}$, forming the superfluid state seen in \textbf{Theory A}. Given how natural the state predicted in \textbf{Theory A} is, one might wonder if the metallic state predicted by \textbf{Theory B} is ever stable to baryon condensation. We leave this question for future work.  

%Although we have not provided any quantitative support, we can be confident that this (relatively minimal) picture must hold, if we assume the Aharony dualities to be valid. This illustrates a rather intriguing scenario in which theories related by these dualities can be used to infer properties of each other's phase diagrams.

%\subsection{Role of particle-hole symmetry}

\subsection{Examples in other fermion-fermion dualities} \label{sec:UN-duality-gapless-examples}

%({\hart We should summarize results here and move bulk of the analysis to an appendix. This section is too long and too dense, given that is fairly redundant with the $SU(2)$ case. -- HG})

Thus far, we have focused our analysis on the dual fermionic theories appearing in the $SU(2)$ quadrality. However, the above considerations are  broadly applicable to any pair of fermionic theories related by a Chern-Simons matter duality. To illustrate this point, we consider a more general composite fermion duality describing a transition between the $\nu=1/k$ Laughlin state and an insulator. This duality relates a theory of Dirac composite fermions with $k-1$ (Abelian) fluxes attached (\textbf{Theory A$'$}) to a theory of composite fermions coupled to a $U(N)$ gauge field (\textbf{Theory B$'$}) %parameterized by the integers $k$ and $N$ 
\cite{Hui2019}, 
\begin{align}
	\mathcal{L}_{A'}(k) = i \bar{\psi}\slashed{D}_a \psi - \frac{1}{2} \frac{1}{4\pi} ada + \frac{1}{2\pi} adv &+  \frac{k-1}{4\pi}vdv + \frac{1}{2\pi}vdA + \dots \, , \label{eqn:theory-c} \\
	\notag &\updownarrow \\
	\mathcal{L}_{B'}(k,N) = i\bar{\eta} \slashed{D}_u \eta - \frac{1}{2}\frac{1}{4\pi} \Tr\left[udu - \frac{2i}{3}u^3\right] & - \frac{N-k}{4\pi} bdb - \frac{1}{2\pi} \Tr[u]db + \frac{1}{2\pi}bdA + \dots  \label{eqn:theory-d} %, \quad a \in U(N) 
\end{align}
In \textbf{Theory A$'$}, $\psi$ is a Dirac fermion charged under an emergent $U(1)$ gauge field $a$, and $v$ is another $U(1)$ gauge field. In \textbf{Theory B$'$}, $\eta$ is a Dirac fermion in the fundamental representation of $U(N)$, $u$ is a $U(N)$ gauge field, and $b$ is a $U(1)$ gauge field. These dualities are derived in Appendix \ref{sec:appendix-UN-duality}, where it is also shown that these theories are dual composite fermion descriptions of the bosonic Landau-Ginzburg theory for the $\nu=1/k$ Laughlin state: when $k$ is even, the fundamental charges are bosons, while when $k$ is odd they are fermions (note that when $N=k=2$, %we can recover \textbf{Theory A} from \textbf{Theory A$'$} by integrating out $c$
we recover \textbf{Theory A} and \textbf{Theory B}, which are associated with the $\nu=1/2$ bosonic Laughlin state). We emphasize that the above duality holds \emph{regardless} of the rank $N$ of the gauge group $U(N)$ in \textbf{Theory B$'$}, and hence amounts to the statement that $\mathcal{L}_{A'}(k)$ is dual to an infinite number of theories $\mathcal{L}_{B'}(k,N)$ parameterized by the integer $N$. %for a given value of $k$, \textbf{Theory A$'$} is dual to \textbf{Theory B$'$}, \emph{regardless} of the rank, $N$, of the gauge group $U(N)$ in the latter theory.

\begin{table}
 \centering
 %\large
 \begin{tabular}{||c || c | c | c | c ||} 
 \hline
  $\nu_*$ & $1/(k-1)$ & $1/(k-2)$ & $3/(3k-4)$ & $2/(2k-3)$ \\
 \hline
 $\nu_\psi+1/2$ & $\infty$ & $2$ & $4$ & $3$ \\ %
 \hline
 $\nu_\eta+1/2$ & $2$ & $\infty$ & $3$ & $4$ \\ 
 \hline
 \textbf{Theory A$'$} & Metal & \makecell{$U(1)_{k-2}$ ($k\neq 2$) \\ Superfluid ($k=2$)} & Jain & Jain \\
 \hline
 \textbf{Theory B$'$} ($N=2)$ & $U(2)_{2,2(k-1)}$ & Metal & $U(3)_{2,3k-4}$ & $U(4)_{2,2(2k-3)}$  \\
 \hline
\end{tabular}
\caption{Fillings at which one of \textbf{Theory A$'$}, Eq. \eqref{eqn:theory-c}, and \textbf{Theory B$'$}, Eq. \eqref{eqn:theory-d}, predicts a metallic ground state and the other a nonmetallic state (first two columns), or where both predict distinct topological orders (last two columns). Here, $N$ is the rank of the $U(N)$ gauge group in \textbf{Theory B$'$}, which is always equal to two in these examples. By Jain we mean a $U(1)^{\mathrm{spin}} \times U(1)$ theory describing the usual Abelian Jain state at filling $\nu_*$.}%{\hart(It also might be worthwhile to have a table here for the gapped cases -- HG)}} 
\label{table: UN metallic fillings}
\end{table}

As in the examples encountered thus far, we find special filling fractions, $\nu_*$, at which the dual theories predict differing ground states.  %such special filling fractions only occur when $N=2$. 
The details of this analysis are essentially identical to that of the preceding $SU(2)$ examples %and so we here only present the results of the analysis and refer the reader 
and are presented in detail in Appendix \ref{sec:appendix-UN-duality-gapless-examples}. Our results are summarized in Table \ref{table: UN metallic fillings}. For example, at the filling $\nu_* = 2/(2k-3)$, \textbf{Theory A$'$} predicts the usual Abelian Jain state, while \textbf{Theory B$'$} predicts the more exotic, non-Abelian $U(4)_{2,2(2k-3)}$ state. Our dynamical proposal for understanding the transitions between these states, as well as the others featured in Table \ref{table: UN metallic fillings}, is essentially identical to that of Section \ref{sec: dynamical scenario}, and so we will not comment on it further. 

One state of particular note is the $U(3)_{2,-1}$ topological order, which is level-rank dual to $U(2)_{-3,-1}^{\mathrm{spin}}$, predicted by \textbf{Theory B$'$} with $k=1$ at $\nu_*=-3$. Now, the non-spin $U(2)_{3,1}$ Chern-Simons theory is dual to the (also non-spin) $(G_2)_1$ Chern-Simons theory \cite{Cordova2019}. Remarkably, $(G_2)_1$ describes precisely the Fibonacci topological order, which supports a single non-trivial anyon, $\tau$, obeying the fusion rule $\tau \times \tau = 1 + \tau$, and is of particular import from the perspective of topological quantum computation \cite{Nayak2008}. It is rather interesting that \textbf{Theory B$'$} predicts the (spin) Fibonacci topological order to appear in competition with the $\nu_*=-3$ IQH state. In future work, we will explore how other mechanisms can lead to the emergence of this exotic order, using bosonic theories, in the spirit of our earlier construction \cite{Goldman2019}.

%The cases in which one theory predicts a metallic state and the other a gapped state are more interesting, having, as we will see, more physical relevance, and so constitute the focus of the remainder of this subsection.
%At fillings $\nu_* = 1/(k-1)$, \textbf{Theory A$'$} predicts a metallic state, whereas \textbf{Theory B$'$} predicts a non-Abelian topological order. These cases parallel those discussed in the previous subsection, in which we found a duality between a paired state of the Abelian theory and an IQH state of the non-Abelian theory.  %Applying our analysis of the $SU(2)$ theories in the preceding subsections to the present problem, our proposal is again that the ground states predicted by both theories are valid phases in the phase diagram at these filling fractions. Indeed, we can again define a dimensionless coupling $\bar{\lambda}_{B'} = g_{\mathrm{YM}}^2 / \omega_c$, where $g_{\mathrm{YM}}^2$ is the Yang-Mills coupling for $u$. We conjecture that the $\bar{\lambda}_{B'} \to \infty$ limit, in which the Yang-Mills term vanishes and the $\eta$ fermions are deconfined, yields the topologically ordered phase in \textbf{Theory B$'$} whereas the $\bar{\lambda}_{B'} \to 0$ limit yields a metallic phase.

%We close this section with the observation that at the filling fractions $\nu_* = 1 / (k-1)$, the topological $U(2)_{2,2(k-1)}$ order predicted by \textbf{Theory B$'$} can be understood as arising from a pairing instability of the metallic composite Fermi liquid state of \textbf{Theory A$'$}. 
%Instead, %we will devote the remainder of this section to 
Additionally, we wish to highlight the cases at fillings $\nu_*=1/(k-1)$, in which the non-Abelian $U(2)_{2,2(k-1)}$ state predicted by \textbf{Theory B$'$} can again be understood as a pairing instability of the Abelian composite Fermi liquid state in \textbf{Theory A$'$}. The argument that $U(2)_{2,2(k-1)}$ can be obtained from pairing in \textbf{Theory A$'$} parallels that for the $U(2)_2$ state discussed in the previous subsection, which corresponds to the special case $k=2$. Indeed, the non-Abelian part of $U(2)_{2,2(k-1)}$ %$ = [SU(2)_2 \times U(1)_{4(k-1)}]/\mathbb{Z}_2$ 
is again $SU(2)_2$, and the major difference from this state lies in the topological spins of the non-Abelian half vortices, which are modified by the level of the Abelian sector. Moreover, one can check explicitly that pairing of the \textbf{Theory A$'$} composite fermions in the $l=2$ angular momentum channel yields the expected $U(2)_{2,2(k-1)}$ topological order \emph{for all $k$} by considering the edge spectrum: this topological order supports three chiral, charge-neutral Majorana fermions and one anti-chiral $U(1)_{k-1}$ bosonic charge mode. From the point of view of \textbf{Theory A$'$}, $l=2$ pairing leads to the three chiral Majorana fermions, in addition to a $U(1)_{k-1}$ charge mode coming from the left over Abelian sector. % the edge theory will therefore consist of a $U(1)_{k-1}$ charge mode, while $l=2$ pairing always yields three chiral Majorana fermions; from this, one can deduce that the bulk anyon spectrum (for $k\neq 1$) is generated by a Majorana fermion, a charge $1/(k-1)$ Laughlin quasiparticle, and a non-Abelian half-vortex, $\sigma_k$, with spin $h_{\sigma_k} = 3/16 + (k-1)/8$. We thus find precisely the expected edge theory and anyon content of the $U(2)_{2,2(k-2)}$ topological order.. %and hence all non-Abelian anyons in this topological order will satisfy the $\mathbb{Z}_2$ fusion rules of the Ising-like twist field, $\sigma$, of $SU(2)_2$. 
%Pairing in the appropriate angular momentum channel likewise ensures that the half-vortices will have the topological spins expected for the anyons of $U(2)_{2,2(k-1)}$. 
That such a large class of non-Abelian topological orders which can be understood via pairing in a composite fermion theory, in this case \textbf{Theory A$'$}, has a dual description as IQH states of composite fermions in a dual non-Abelian theory, \textbf{Theory B$'$} is quite remarkable. %Below, we discuss some examples which are of physical relevance.

We close this section by commenting on some particular examples of physical interest. First, for $k=1$ and $\nu_*=\infty$, \textbf{Theory A$'$} is a free Dirac fermion at finite chemical potential but vanishing magnetic field. In this case, in \textbf{Theory B$'$}, it is possible to integrate out the auxiliary gauge field $b$ without violating flux quantization: its Chern-Simons level is $-(N-k)=-1$. This cancels the Chern-Simons term for the Abelian part of the $U(2)$ gauge field, $\Tr[u]$, Higgsing the background EM field, $A_\mu$, and leading to a chiral superconductor with topological order is $U(2)_{2,0} = [SU(2)_2 \times U(1)_0]/\mathbb{Z}_2 \cong SO(3)_1$, which is \emph{Abelian}. This state contains only a single anyon, a Majorana fermion\footnote{This can be viewed as a consequence of the condensation of the Majorana fermion of the $SU(2)_2$ topological order, which confines the non-Abelian ``half-vortex" of the $SU(2)_2$ factor.} \cite{Seiberg2016,Aharony2017}. Such a state can be accessed via a pairing instability in \textbf{Theory A$'$}. It is natural to expect this instability to arise due to the effects of local four-fermion operators that destabilize \textbf{Theory A$'$} in the order of limits in which the $\eta$ fermions of \textbf{Theory B$'$} form an IQH state. %To see this explicitly, we note that since the $U(1)$ factor has a vanishing Chern-Simons level, the $\mathbb{Z}_2$ quotient simply projects out all $SU(2)_2$ Wilson lines which have non-trivial braiding with the Wilson line transforming in the spin-$1$ representation of $SU(2)_2$, which generates the $\mathbb{Z}_2$ one-form symmetry.\footnote{We recall that $SU(2)_2$ possesses three Wilson lines, transforming in the spin-$0$, spin-$1/2$, and spin-$1$ representations of $SU(2)$. The correspond, respectively, to the vacuum, the non-Abelian Ising-like anyon $\sigma$, and a Majorana fermion $\psi$ (not to be confused with the fermion field $\psi$ of \textbf{Theory A$'$}), respectively. Here, $\psi$ generates a $\mathbb{Z}_2$ one-form symmetry, as $\psi^2 = 1$ and braiding $\psi$ around $\sigma$ yields a phase of negative unity.} In other words, we condense the Majorana fermion of the $SU(2)_2$ topological order, which confines the non-Abelian $\sigma$ anyon of $SU(2)_2$. We are thus left with an Abelian order containing only a single nontrivial anyon, a Majorana fermion, and which therefore describes a chiral topological superconductor. 
%It is reassuring that the dualities lead to such a state in this case and not a non-Abelian topological order, which would be difficult to motivate in the phase diagram of a free Dirac fermion. %Since this theory is free, there is certainly no room for a non-Abelian gapped phase to appear, even were one to introduce Coulomb interactions. Consequently, we predict that the non-Abelian, $U(2)_2\times U(1)_1$ state predicted by \textbf{Theory B$'$} has no basin of attraction, flowing immediately to strong coupling. 

For $k=-1$ and $\nu_*=-1/2$, \textbf{Theory A$'$} is Son's Dirac composite Fermi liquid theory of the half-filled Landau level \cite{Son2015}. In this case, the topological order predicted by \textbf{Theory B$'$} is $U(2)_{2,-4} = [SU(2)_{2}\times U(1)_{-8}]/\mathbb{Z}_2$, which is the (time-reversed%\footnote{The anti-Pfaffian at $\nu=1/2$ is described by $U(2)_{-2,4}$ \cite{Lian2018}, and so it is natural to find the time-reverse conjugated order, $U(2)_{2,-4}$, at $\nu=-1/2$.}
) topological order of the famous anti-Pfaffian state, another paired state of composite fermions \cite{Lee2007,Levin2007}! %This state can be understood as a paired phase of the composite fermions of \textbf{Theory A$'$}. %This representation of the anti-Pfaffian state as an IQH state of composite fermions only has precedent in parton constructions, which hinge on dynamical assumptions that the use of non-Abelian composite fermion dualities does not require.
Additionally, for $k=3$ and $\nu_*=1/2$, \textbf{Theory B$'$} predicts the $[SU(2)_{2} \times U(1)_{8}]/\mathbb{Z}_2 = U(2)_{2,4}$ order, which is that of Wen's $(221)$ parton state \cite{Wen1991,Wen1999}, another proposed ground state of fermions at half filling that can be understood in terms of pairing in \textbf{Theory A$'$}, which now has additional attached fluxes. Now, although these states are all accessible within parton constructions (see, for example, Ref. \cite{Ma2020}), we must again emphasize that the projective framework hinges on the dynamical assumption that the electron fractionalizes into partons which do not confine. Finally, another non-trivial bosonic example is $k=0$ and $\nu_*=1$, in which \textbf{Theory B$'$} predicts a $[SU(2)_{2} \times U(1)_{-4}]/\mathbb{Z}_2 = U(2)_{2,-2}$ ground state, which describes the Ising topological order \cite{Seiberg2016a}. While these results are reminiscent of the parton constructions giving these states, the use of duality provides a connection between partonic intuition and the dynamics of pairing. Additionally, it is straightforward to check in these examples that $l=2$ pairing in \textbf{Theory A$'$} yields the expected $U(2)_{2,2(k-1)}$ order by comparing the edge theories. %In contrast, our use of non-Abelian composite fermions supplies a deconfining non-Abelian gauge field, putting the dynamical stability of these non-Abelian topological orders on a firmer footing. 
It would be interesting to determine if other dual descriptions exist which yield the other proposed Pfaffian-like states, which arise from pairing the composite fermions of \textbf{Theory A$'$} in other angular momentum channels, as IQH phases of dual composite fermions, and we leave this to future work.

\section{Building non-Abelian states from excitonic pairing}
\label{sec:pairing}

In the preceding section, we illustrated how non-Abelian Chern-Simons matter dualities may be used to map out parts of the phase diagram for electrons (or bosons) in a magnetic field at certain fractional fillings, $\nu_*$, finding gapless states as well as both Abelian and non-Abelian topological orders. %Although this constitutes the main message of this paper, we would now like 
We now turn to present an alternative means of constructing non-Abelian FQH states, still making use of the dual composite fermion theories employed above. Our goal here is to provide a complementary perspective to our previous work \cite{Goldman2019}, in which we constructed Landau-Ginzburg theories for the Read-Rezayi states using bosonic Chern-Simons-matter using a multilayer pairing procedure. %Moreover, by employing fermionic theories which are dual to the conventional Landau-Ginzburg theories for FQH states, we provide our following construction with a stronger microscopic grounding than that found in parton constructions of non-Abelian states, which require fractionalization of the electron by hand and deconfinement of the resulting partons as an assumption.
Specifically, we will consider condensing interlayer excitons (pairs of fermions in different layers which are neutral under the external EM gauge field) in \textbf{Theory B}, Eq. \eqref{eqn:TheoryD}, to generate non-Abelian states. We will find, however, that the excitonic paired phases are not the Read-Rezayi states constructed in Ref. \cite{Goldman2019}, but rather the Blok-Wen states with $U(k)_2$ topological order, as in Section \ref{sec:fermions}. 

For simplicity, we again consider a bilayer system, with each layer being a $\nu=1/2$ bosonic Laughlin state. We use the dual fermionic description of \textbf{Theory B} for each layer so that the (initially decoupled) multilayer system is described by the Lagrangian
\begin{align}
	\mathcal{L}_{B,2} = \sum_{n=1}^2 \left( i\bar\chi_n\slashed{D}_{b_n- A \mathbf{1}/2}\chi_n - \frac{1}{2} \frac{1}{4\pi} \mathrm{Tr}\left[ b_ndb_n - \frac{2i}{3} b_n^3 \right] \right) - \frac{1}{2} \frac{1}{4\pi} AdA.
\end{align}
Here, $\chi_n$ and $b_n$ are the composite fermions and $SU(2)$ gauge fields on layer $n$, respectively. Note that each layer couples in the same way to the external electromagnetic field, $A$. Moreover, we see from Table \ref{tab:shared-IQH-states} that when each layer is at filling $\nu =  1/2$, so that the full multilayer system is at filling $\nu = 1$, the $\chi_n$ of each layer form an IQH state.

Now, %akin to our previous bosonic construction \cite{Goldman2019}
following the standard approach \cite{Jackiw1981}, we introduce an interlayer excitonic pairing interaction mediated by an electrically neutral scalar field, $\Sigma$,
\begin{align}
	\mathcal{L}_{\mathrm{exciton}} = \bar{\chi}_1 \Sigma \chi_2 + \mathrm{H.c.}
\end{align}
The field $\Sigma$ can be thought of as a Hubbard-Stratonovich field for the pairing interaction. It couples minimally to the gauge fields on either layer, and so has dynamics described by
\begin{align}
	\mathcal{L}_\Sigma = |\partial \Sigma -ib_1 \Sigma + i \Sigma b_2|^2 - V[\Sigma], \label{eqn:Sigma-lagrangian}
\end{align}
where $V[\Sigma]$ is the potential for $\Sigma$. Under gauge transformations,
\begin{align}
	\Sigma \mapsto U_1 \Sigma U_2^\dagger, \qquad U_m \in SU(2) \text{ on layer } m. \label{eqn:Sigma-gauge-transformation}
\end{align}

The full multi-layer theory is therefore described by the Lagrangian,
\begin{align}
	\mathcal{L}_{\mathrm{bilayer}} = \mathcal{L}_{B,2} + \mathcal{L}_{\mathrm{exciton}} + \mathcal{L}_\Sigma \, .
\end{align}
The condensation of $\Sigma$ yields the excitonic paired state, characterized by a non-zero expectation value for the operator $\bar{\chi}_1 \chi_2$. %Here  we view $\mathcal{L}_{\mathrm{bilayer}}$ as an effective action describing the transition to the excitonic paired state, although $\Sigma$ should in principle arise as a Hubbard-Stratonovich field in the decoupling of some interaction involving the $\chi_n$ fermions. 
It should be emphasized that in the dual descriptions of \textbf{Theory A} and the Landau-Ginzburg theory of Eq. \eqref{eq: U(1)2}, the interaction $\mathcal{L}_{\mathrm{exciton}}$ will correspond to a highly nonlocal object involving monopole operators. In general, the fundamental fields do not map to local operators under the dualities of Eq. \eqref{eq: U/SU}-\eqref{eq: U/U} \cite{Aharony2016}. Hence, the upshot of examining this dual fermionic theory is that we can access regions of the phase diagram at a given filling fraction, which are less readily understood in the formulation of \textbf{Theory A} or the original bosonic Landau-Ginzburg theory, as they would require the inclusion of complicated, nonlocal interactions.

Now, suppose we have a nonzero magnetic field such that each layer is at filling $\nu = 1/2$, meaning the $\chi_n$ each fill a single Landau level, as indicated in Table \ref{tab:shared-IQH-states}. We assume that we can safely integrate out the occupied $\chi_n$ Landau levels, yielding additional level $-1/2$ Chern-Simons terms for the $b_m$ gauge fields. The effective action describing %fluctuations about 
this gapped state is then %essentially takes the same form as $\mathcal{L}_{\mathrm{bilayer}}$ above, but with $SU(2)_{-1}$ Chern-Simons terms for the dynamical gauge fields $b_m$ (and a  $U(1)_{-1}$ Chern-Simons term for the external probe field, $A$). 
\begin{align}
	\mathcal{L}_{\mathrm{bilayer}} = -\sum_{n=1}^2  \frac{1}{4\pi} \mathrm{Tr}\left[ b_ndb_n - \frac{2i}{3} b_n^3 \right]  - \frac{1}{4\pi} AdA + \tilde{\mathcal{L}}_\Sigma+\dots \, ,
\end{align}
%where the $\chi_n$ fields now describe holes in the occupied Landau level. %\textcolor{violet}{(I'm not sure of the appropriate way to write down the effective action for the fermions, when expanding about the filled LL state. Do we need to add a mass and chemical potential?)}. %\footnote{\textcolor{blue}{To alleviate the notation, here we do not make explicit the fermion mass and  the chemical potential  needed to tune the theory to a filled Landau level.}} 
where we have integrated out the fermions $\chi$, leading to a renormalized Lagrangian for $\Sigma$, denoted $\tilde{\mathcal{L}}_\Sigma$. %Note that the Chern-Simons terms for the $b_n$ fields are now both at level $-1$. 
If the potential $V[\Sigma]$ is such that the field $\Sigma$ is massive and does not condense, then we are simply left with an $SU(2)_{-1}^{\mathrm{spin}} \times SU(2)_{-1}^{\mathrm{spin}} \leftrightarrow U(1)_2 \times U(1)_2$ Chern-Simons theory at low energy, describing two decoupled layers of $\nu = 1/2$ Laughlin states. Now, suppose instead that the potential $V[\Sigma]$ is such that $\Sigma$ obtains a non-zero vacuum expectation value, $\langle  \Sigma \rangle \propto \mathbf{1}$.
In this excitonic paired state, the gauge group $SU(2) \times SU(2)$ will be Higgsed down to the diagonal $SU(2)$ subgroup, as follows from the gauge transformations of Eq. \eqref{eqn:Sigma-gauge-transformation}. Explicitly, from the Lagrangian for $\Sigma$, the linear combination $b_1 - b_2$ of the gauge fields acquires a mass, effectively identifying the gauge fields of each layer: $b_1 \equiv b_2 \equiv b$. Hence, the Chern-Simons terms will add, resulting in a non-Abelian $SU(2)_{-2}^{\mathrm{spin}}\leftrightarrow U(2)_2$ Chern-Simons theory at low energies. %\textcolor{violet}{(I am still a little uncertain now about how reasonable it is to have pairing on top of a Landau level.)}  %RS: Is the preceding discussion too schematic?

%Therefore, excitonic pairing in the language of \textbf{Theory B} yields a non-Abelian topological order. %\textcolor{violet}{(RS: The following issue of spin gauge fields/invisible fermionic lines should first be addressed in Section \ref{sec:fermions}.)} 
%However, one must be careful in identifying the anyon content of this topological order since, as in the discussion in Section \ref{sec:fermions} \textcolor{violet}{(RS: to be added, if necessary)}, the gauge fields $b_m$ appearing above must be spin gauge fields, being coupled to fermions. One can use the level-rank dualities to re-express the theory in terms of non-spin gauge fields \cite{Ma2020} : $SU(2)_{-2} \leftrightarrow U(2)_2$. Hence, the excitonic paired state corresponds to a $U(2)_2$ topologically ordered state. \textcolor{blue}{EF: As we need to be carefully l with the jargon that we use and be clear that we need a spin$_c$ connection.}

%\textcolor{violet}{(RS: The following needs some cleaning up, but I wanted to check if we're all on the same page about this first.)} 
It is possible to explicitly write out the anyon spectrum of this $U(2)_2$ topological order in terms of composite operators of the fundamental fermions. This requires identifying the operators which transform under the non-trivial spin-$1/2$ and spin-$1$ representations of the $SU(2)_{-2}$ gauge theory and taking into account additional spin factors coming from the underlying fermionic statistics of said operators. For reference, we denote the anyons transforming in the spin-$1/2$ and spin-$1$ representations of the $SU(2)_{-2}$ topological order as $\sigma$ and $\psi$ (not to be confused with the fermion field in \textbf{Theory A}), corresponding to the non-Abelian half-vortex (i.e. the Ising twist field) and Abelian Majorana fermion, respectively, in the time-reversed conjugate of the bosonic $\nu=1$ Moore-Read state. They have spin\footnote{Here, the spin of an anyon $a$ is the phase factor $\exp(2\pi i h_a)$ picked up when rotating it through an angle of $2\pi$. This is not to be confused with the spin-$j/2$ representations of $SU(2)$.} $h_\sigma = -3/16$ and $h_\psi = 1/2$, respectively, and satisfy the fusion rules $\sigma \times \sigma = 1 + \psi$ and $\psi \times \psi = 1$, where $1$ represents the vacuum. Now, in our theory, the minimal charge anyons are represented by the fermions $\chi_1$, which transform in the fundamental (or spin-$1/2$) representation of $SU(2)_{-2}$. The $\chi_1$ operators will thus satisfy the same fusion rules as the $\sigma$ anyon in the $SU(2)_{-2}$ theory, but they have spin $-3/16+1/2 = 5/16$, due to the bare fermionic statistics of $\chi_1$. The other non-trivial anyon is represented by the composite operator $\bar{\chi}_1 \tau^a \chi_1$, where $\tau^a$ is the vector of generators of $SU(2)$. This operator is charge neutral and transforms in the spin-$1$ representation of $SU(2)$, meaning it obeys the same fusion rules as $\psi$. It also has the same spin as $\psi$, $h_\psi = 1/2$, since it has bare bosonic statistics, being a bilinear in fermion operators. Once can check that these anyons with these spins match the anyon spectrum expected for the $U(2)_2$ topological order. % \textcolor{violet}{(RS: see below)}. 
Finally, note that the fundamental fermions $\chi_1$ and $\chi_2$ are indistinguishable in the excitonic paired phase, as one can be transmuted into the other via the the $\langle \bar{\chi}_1 \chi_2 \rangle$ condensate. Hence, there is no double-counting of anyons.

Several remarks on this construction are in order. This excitonic pairing mechanism is somewhat unconventional and differs from the more common Read-Green construction \cite{Read2000} used to describe the Moore-Read states. In the latter picture, the electrons (or bosons) are mapped to composite fermions using non-relativistic flux attachment. At the appropriate filling fractions, the composite fermions see an effectively vanishing flux at mean-field level. The resulting composite Fermi liquid can give way to a pairing instability in the $p+ip$ channel, Higgsing the dynamical $U(1)$ Chern-Simons gauge field down to its $\mathbb{Z}_2$ subgroup and resulting in a gapped state. The non-Abelian Ising anyons in the Moore-Read state then have a description in terms of vortices of the Chern-Simons gauge field. In the present construction, we are instead pairing fermions on top of a filled Landau level, a gapped state. Hence, unlike the Read-Green picture, we cannot understand our exciton paired state as arising from some perturbative instability, since interactions must be sufficiently strong to overcome the gap. In addition, one can check from standard homotopy arguments that the symmetry breaking pattern $SU(2) \times SU(2) \to SU(2)_{\mathrm{diagonal}}$ does not admit vortex configurations \cite{Fradkin1999,Goldman2019}. Instead, the anyon spectrum in our model is generated by composite objects formed from the fundamental fermions, as outlined above. %which transform as nontrivial representations of the residual diagonal $SU(2)$ gauge group.  \textcolor{blue}{EF: perhaps we can write this down more explicitly?} 
It should be noted that even our earlier bosonic construction \cite{Goldman2019} required a similarly unconventional pairing mechanism, in which it was necessary to assume that composite bosons paired rather than condensed. Finally, it is clear that we can generalize our construction to a multilayer system with $k$ copies of the $\nu=1/2$ Laughlin state; interlayer excitonic pairing in such a system would lead to a $U(k)_2$ topological order.

\section{Discussion}

Employing non-Abelian composite fermion dualities, we have presented two complementary pictures for describing a broad range of non-Abelian FQH states, which can be obtained either as IQH states of non-Abelian composite fermions or as excitonic states in multilayer systems. %, including the $U(k)_2$ Blok-Wen states, 
%while also uncovering how these dualities provide new insight into the dynamics of the theories they relate. This builds on our earlier construction \cite{Goldman2019} in which we made use of dual non-Abelian bosonic theories to develop Landau-Ginzburg theories for the Read-Rezayi and generalized non-Abelian spin singlet states. 
Along the way, we developed new insights into the non-Abelian theories' dynamics, in which the order of the lowest Landau level ($B\rightarrow\infty$) and IR limits was seen to play a crucial role in determining the ultimate choice between the non-Abelian ground state and a competing Abelian state that is natural in a dual description. This subtlety has thus far received little attention in studies of non-Abelian dualities, yet we find it to be a ubiquitous feature of non-Abelian fermion-fermion dualities. It may be a worthwhile endeavour to see whether studying these theories at finite magnetic field in the `t Hooft limit, in the vein of Ref. \cite{Halder2019}, may provide an analytical handle on the physics of these transitions -- we leave this for future work.  %In the present work, we first showed that non-Abelian $U(k)_2$ FQH states can be obtained by simply filling up Landau levels of composite fermions in a non-Abelian Chern-Simons-matter theory dual to the usual bosonic Landau-Ginzburg theory for the $\nu=1/2$ Laughlin state. An important consequence of our analysis was that two dual fermionic theories can predict \emph{distinct} ground states (either metallic, Abelian, or non-Abelian states) at the same filling fraction. This constitutes one of our main messages: the dualities can be used to map out regions of the phase diagram of electrons at fractional filling. Specifically, we argued that it is the order in which one takes IR and lowest Landau level limits which selects the physical ground state from the two states predicted by the dual theories. 
Interestingly, related physics has been observed recently in numerics, where it has been argued that the ground state at certain fillings can exhibit effectively Abelian topological order for short-range interactions and non-Abelian order as the interaction range is increased \cite{Yang2019,Andrews2020}. We hope that our work will motivate more numerical efforts in this direction.

Although we cannot make many concrete statements about the transitions we propose to occur between the Abelian and non-Abelian states, we remarkably find several examples in which the non-Abelian states -- among them the anti-Pfaffian -- can be understood in terms of pairing in a dual composite Fermi liquid description. Such dualities between composite fermion pairing and the IQH effect in a dual, non-Abelian theory are new, and finding new examples of such dualities will be a fruitful direction for future work. Looking forward, a natural question to ask is whether non-Abelian fermion-fermion dualities can be used to derive other non-Abelian FQH states, beyond the variations of the Blok-Wen states we find. Indeed, although the anti-Pfaffian is a member of the $U(2)_{2,2(k-1)}$ series of states, we do not seem to arrive at the Pfaffian or PH-Pfaffian states. To that end, it may be fruitful to apply our analysis using dualities involving Chern-Simons-matter theories with gauge groups other than $SU(N)$ or $U(N)$, as the family of Pfaffian states can naturally be described using $O(2)_{2,L}$ Chern-Simons theories \cite{Cordova2018a}. It is also somewhat peculiar in that the Read-Rezayi and generalized non-Abelian spin singlet states, which are readily obtained through non-Abelian bosonic theories, do not appear to be accessible within the present approach. We leave the construction of fermionic theories for these states to future work. Conversely, we found that the (spin) Fibonacci topological order appears naturally in Sec. \ref{sec:UN-duality-gapless-examples} within this composite fermion approach. It would be interesting to see whether a variation of our earlier bosonic construction \cite{Goldman2019} may allow for accessing this exotic state as well. %among which is the anti-Pfaffian, appear to be naturally described as pairing instabilities of a composite Fermi liquid of the dual Abelian theory. %As a complement to this first line of attack in constructing non-Abelian FQH states, we subsequently provided a construction closer in spirit to Ref. \cite{Goldman2019} in which, by stacking $k$ copies of $\nu=1/2$ Laughlin states and introducing interlayer excitonic pairing in their dual fermionic descriptions, one arrives again at the $U(k)_2$ Blok-Wen states.

\section*{Acknowledgements}
We thank Gil Young Cho, Michael Mulligan, and Chong Wang for useful discussions. This work was supported in part by the National Science Foundation under Grant No. DMR-1725401 at the University of Illinois (EF, HG, RS). RS also acknowledges the support of the Natural Sciences and Engineering Research Council of Canada (NSERC) [funding reference number 6799-516762-2018]. HG also acknowledges support from 
the NSF Graduate Research Fellowship Program under Grant No. DGE-1144245 and from the Gordon and Betty Moore Foundation EPiQS Initiative through Grant No. GBMF8684 %({\hart get grant no.}) 
at the Massachusetts Institute of Technology.

\begin{appendix}
\section{Chern-Simons Conventions \label{sec:appendix-conventions}}
In this appendix, we lay out our conventions for non-Abelian Chern-Simons gauge theories. We define $U(N)$ gauge fields $a_\mu=a^b_\mu t^b$, where $t^b$ are the (Hermitian) generators of the Lie algebra of $U(N)$, which satisfy $[t^a,t^b]=if^{abc}t^c$, where $f^{abc}$ are the structure constants of $U(N)$. The generators are normalized so that $\Tr[t^bt^c]=\frac{1}{2}\delta^{bc}$. The trace of $a$ is a $U(1)$ gauge field, which we require to satisfy the Dirac quantization condition,
\be
\label{eq: flux quantization}
\int_{\Sigma} \frac{d\Tr[a]}{2\pi}=n\in\mathbb{Z}\,.
\ee
where $\Sigma\subset X$ is an oriented 2-cycle in spacetime, which we denote $X$. If $a_\mu$ couples to fermions, then it is a spin$_c$ connection, and it satisfies a modified flux quantization condition
\be
\int_\Sigma\frac{d\Tr[a]}{2\pi}=\int_\Sigma\frac{w_2}{2}+n\,,\,n\in\mathbb{Z}\,,
\ee
where $w_2$ is the second Stiefel-Whitney class of $X$. In general, the Chern-Simons levels for the $SU(N)$ and $U(1)$ components of $a$ can be different. We therefore adopt the standard notation \cite{Aharony2016},
\be
U(N)_{k,k'}=\frac{SU(N)_k\times U(1)_{Nk'}}{\mathbb{Z}_N}\,.
\ee
By taking the quotient with $\mathbb{Z}_N$, we are restricting the difference of the $SU(N)$ and $U(1)$ levels to be an integer multiple of $N$,
\be
k'=k+nN\,,n\in\mathbb{Z}\,.
\ee
This enables us to glue the $U(1)$ and $SU(N)$ gauge fields together to form a gauge invariant theory of a single $U(N)$ gauge field $a=a_{SU(N)}+\tilde{a}\,\mathbf{1}$, with $\Tr[a]=N\tilde{a}$ having quantized fluxes as in Eq. \eqref{eq: flux quantization}. The Lagrangian for the $U(N)_{k,k'}$ theory can be written as
\be
\mathcal{L}_{U(N)_{k,k'}}=\frac{k}{4\pi}\Tr\left[a_{SU(N)}da_{SU(N)}-\frac{2i}{3}a_{SU(N)}^3\right]+\frac{Nk'}{4\pi}\tilde{a}d\tilde{a}\,.
\ee
For the case $k=k'$, we simply refer to the theory as $U(N)_k$. 

Throughout this paper, we implicitly regulate non-Abelian (Abelian) gauge theories using Yang-Mills (Maxwell) terms, as opposed to dimensional regularization \cite{Witten1989,Chen1992}. In Yang-Mills regularization, there is a one-loop exact shift of the $SU(N)$  level, $k\rightarrow k+\operatorname{sgn}(k)N$, that does not appear in dimensional regularization. Consequently, to describe the same theory in dimensional regularization, one must start with a $SU(N)$ level $k_{\mathrm{DR}}=k+\operatorname{sgn}(k)N$. The dualities discussed in this paper, e.g. Eqs. \eqref{eq: U/SU}-\eqref{eq: U/U}, therefore would take a somewhat different form in dimensional regularization.

\section{Details of $U(N)$ Fermion-Fermion Duality Examples}

In this Appendix we provide the details of the analysis outlined in Section \ref{sec:UN-duality-gapless-examples}. We begin by deriving the duality between Eq. \eqref{eqn:theory-c} and Eq. \eqref{eqn:theory-d} and then identify the states listed in Table \ref{table: UN metallic fillings}.

\subsection{Derivation of the Duality \label{sec:appendix-UN-duality}}

As noted in Section \ref{sec:UN-duality-gapless-examples}, the fermionic theories \textbf{Theory A$'$} and \textbf{Theory B$'$} are both dual to the bosonic Landau-Ginzburg theory for the $\nu=1/k$ Laughlin state, which is described by the Lagrangian
\begin{align}
	\mathcal{L}_{\Phi}(k) = |D_b\Phi|^2 - |\Phi|^4 + \frac{k}{4\pi}bdb + \frac{1}{2\pi}bdA .
\end{align}
Here, $\Phi$ is a complex scalar field, $b$ is a $U(1)$ gauge field, and $A$ is the external electromagnetic field. It is straightforward to see that one obtains the Laughlin state when $\Phi$ is gapped by a mass and a trivial insulator when $\Phi$ condenses.

In order to derive these dualities, we take as our starting point the $SU/U$ duality of Eq. \eqref{eq: SU/U},
\begin{align}
	|D_A\Phi|^2 - |\Phi|^4 \longleftrightarrow i\bar{\eta} \slashed{D}_u \eta - \frac{1/2}{4\pi} \Tr\left[udu - \frac{2i}{3}u^3\right] - \frac{1}{2\pi} \Tr[u]dA - \frac{N}{4\pi}AdA, \quad a \in U(N),
\end{align}
where $u$ is a $U(N)$ gauge field and $\eta$ is a fermion in the fundamental representation of $U(N)$. Note that the rank $N$ can be an \emph{arbitrary} integer, and so the above equation implies that the Wilson-Fisher theory is dual to an infinite number of fermionic $U(N)$ gauge theories. Now, one can derive new dualities from old ones by applying the modular transformations \cite{Witten2003}
\be
\mathcal{S}: \mathcal{L}[A]\mapsto\mathcal{L}[b]+\frac{1}{2\pi}Adb\,,\qquad\mathcal{T}:\mathcal{L}[A]\mapsto\mathcal{L}[A]+\frac{1}{4\pi}AdA\,,
\ee
to both sides of a duality, where again $A$ is the background EM field and $b$ is a new dynamical $U(1)$ gauge field. Here, $\mathcal{S}$ is the operation of promoting a background gauge field to a dynamical one, while $\mathcal{T}$ corresponds to the addition of a Landau level. Applying $\mathcal{S} \mathcal{T}^k$ to the $SU/U$ duality yields,
\begin{align}
	\mathcal{L}_{\Phi}(k) \longleftrightarrow i\bar{\eta} \slashed{D}_u \eta - \frac{1/2}{4\pi} \Tr\left[udu - \frac{2i}{3}u^3\right] - \frac{1}{2\pi} \Tr[u]db - \frac{N-k}{4\pi}bdb + \frac{1}{2\pi}bdA = \mathcal{L}_{B'}(k,N) \, .
\end{align}
On the other hand, we can also consider the Abelian bosonization duality \cite{Karch2016,Seiberg2016},
\begin{align}
	|D_A \Phi|^2 - |\Phi|^4 \longleftrightarrow i \bar{\psi}\slashed{D}_a \psi - \frac{1}{2} \frac{1}{4\pi} ada + \frac{1}{2\pi} adA - \frac{1}{4\pi}AdA \, ,
\end{align}
where $a$ is an emergent $U(1)$ gauge field and $\psi$ a Dirac fermion. Applying $\mathcal{S} \mathcal{T}^k$ to this duality, we find
\begin{align}
	\mathcal{L}_{\Phi}(k) \longleftrightarrow i \bar{\psi}\slashed{D}_a \psi - \frac{1}{2} \frac{1}{4\pi} ada + \frac{1}{2\pi} adv + \frac{k-1}{4\pi}vdv + \frac{1}{2\pi}vdA = \mathcal{L}_{A'}(k) .
\end{align}
We thus arrive at the desired dualities
\begin{align}
	\mathcal{L}_{A'}(k) \longleftrightarrow \mathcal{L}_{\Phi}(k) \longleftrightarrow \mathcal{L}_{B'}(k,N) .
\end{align}
%where
%\begin{align}
%	\mathcal{L}_C &= i \bar{\psi}\slashed{D}_b \psi - \frac{1}{2} \frac{1}{4\pi} bdb + \frac{1}{2\pi} bdc + \frac{k-1}{4\pi}cdc + \frac{1}{2\pi}cdA
%\end{align}
%and
%\begin{align}
%	\mathcal{L}_D &= i\bar{\eta} \slashed{D}_a \eta - \frac{1/2}{4\pi} \Tr\left[ada - \frac{2i}{3}a^3\right] - \frac{N-k}{4\pi}cdc - \frac{1}{2\pi} \Tr[a]dc + \frac{1}{2\pi}cdA, \quad a \in U(N),
%\end{align}
We emphasize that these dualities hold true for any value of the rank, $N>0$, of the gauge group $U(N)$ of \textbf{Theory B$'$}.

\subsection{Examples involving gapless states \label{sec:appendix-UN-duality-gapless-examples}}

Let us now investigate the states predicted by the dual theories, \textbf{Theory A$'$} and \textbf{Theory B$'$}, at fractional electronic filling fractions, following the logic in our study of the dual fermionic theories in the $SU(2)$ quadrality in Section \ref{sec:fermions}. We define the filling fraction of the $\psi$  composite fermions as
\begin{align}
	\nu_\psi = \frac{2\pi\langle \psi^\dagger \psi \rangle}{\langle \varepsilon^{ij} \partial_i a_j \rangle}.
\end{align}
Using the equations of motion of $\mathcal{L}_{A'}$, we find the following relationship between the electronic and $\psi$ filling fractions
\begin{align}
	\nu_\psi = \frac{1}{2} + \frac{1}{-1/\nu + (k-1)} \iff \nu = \frac{2\nu_\psi-1}{2(k-1)\nu_\psi - k - 1}. \label{eqn:nu-psi}
\end{align}
As for the composite fermions of the non-Abelian \textbf{Theory B$'$}, to define the $\eta$ filling fraction, we first decompose the $U(N)$ gauge field as $u = u_{SU(N)} + \tilde{u}\mathbf{1}$, where $\mathbf{1}$ is the $N\times N$ identity matrix, $\tilde{u}$ is a $U(1)$ gauge field, and $u_{SU(N)}$ is an $SU(N)$ gauge field. In the presence of a non-zero $U(1)$ flux, $\langle \varepsilon^{ij} \partial_i \tilde{u}_j \rangle$, the $\eta$ fermion Landau level degeneracy is given by
\begin{align}
	d_{LL} = \frac{\langle \varepsilon^{ij} \partial_i \tilde{u}_j \rangle A}{2\pi} \times \text{(color degeneracy)} \times \text{(charge)} =  \frac{\langle \varepsilon^{ij} \partial_i \tilde{u}_j \rangle A}{2\pi}  \times N \times 1 = \frac{N\langle \varepsilon^{ij} \partial_i \tilde{u}_j \rangle A}{2\pi}.
\end{align}
Hence, the $\eta$ fermion filling fraction is given by\footnote{We add the minus sign for consistency with the definition of the $\chi$ filling fraction $\nu_\chi$ of Eq. \eqref{eqn:nu_chi} when $N=k=2$.}
\begin{align}
	\nu_\eta = -\frac{2\pi \langle \eta^\dagger \eta \rangle}{N \langle \varepsilon^{ij} \partial_i \tilde{u}_j \rangle}.
\end{align}
Using the equations of motion of $\mathcal{L}_{B'}$, we find
\begin{align}
	\nu_\eta = -\frac{1}{2} + \frac{N}{1/\nu + N-k} \iff \nu =  \frac{2\nu_\eta + 1}{2(k-N)\nu_\eta + (k+N)}. \label{eqn:nu-eta}
\end{align}
Let us suppose the $\psi$ fermions fill up an integer number of LLs, so that $\nu_\psi = p - 1/2$. Then, from Eq. \eqref{eqn:nu-psi}, we have
\begin{align}
	\nu = \frac{p-1}{(p-1)(k-1) - 1}, \label{eqn:TheoryA'-IQH-filling}
\end{align}
which is simply the Jain sequence of states.

We are interested in seeing whether a gapped state of the $\psi$ fermions ever corresponds to a metallic state of the $\eta$ fermions (i.e. with $\nu_\eta \to \infty$). From Eq. \eqref{eqn:nu-eta}, we see that the $\eta$ fermions form a metallic state when	$1/\nu = -N+k \in \mathbb{Z}$. We must therefore look for solutions of the equation
\begin{align}
	k-1 - (p-1)^{-1} = -N+k.
\end{align}
The only valid solution with $N>0$ is $(N,p) = (2,2)$. So, when $N=2$ and the $\psi$ fermions fill the $p=2$ Landau level, the $\eta$ fermions form a metallic state. The electronic filling fraction is $\nu_* = 1/(k-2)$. At this filling, we can integrate out the $\psi$ fermions in \textbf{Theory A$'$} to obtain the effective action
\begin{align}
	\mathcal{L}_{A',\mathrm{eff}} &= \frac{k-2}{4\pi} vdv + \frac{1}{2\pi} vdA \, .
\end{align}
We thus have two cases to consider: $k=2$ and $k \neq 2$. When $k=2$, \textbf{Theory A$'$} yields the usual dual theory for a superfluid (recall that there is an implicit Maxwell term in the action for $c$). This is not surprising, as the filling fraction of the electrons (which are bosons for $k=2$) is $\nu = \infty$, which is to say they see no magnetic field. Hence, for $k=2$, both \textbf{Theory A$'$} and \textbf{Theory B$'$} predict compressible states. %It is thus possible that they both describe the \emph{same} phase, in contrast to the situation described in the previous subsection for the dual fermion theories appearing in the $SU(2)$ quadrality. \textcolor{blue}{EF: this argument is problematic. The superfluid and the metal are really different states of matter. The superfluid has a stiffness which yields a finite propagation speed to its Goldstone (phase) mode. This physics is encoded in the local (dual) Maxwell action. However, the effective action of the metal is inherently non-local, it is dissipative, and exhibits Landau damping. So here too the only consistent scenario is not an equivalency but a quantum phase transition of the metal-superfluid variety.} {\hart HG: I agree. See discussion in Section 3.4.}
In contrast, for $k \neq 2$, \textbf{Theory A$'$} describes the incompressible $\nu = 1/(k-2)$ Laughlin state, while \textbf{Theory B$'$} again describes a metallic state of the $\eta$ fermions. %Here it thus appears necessary to make recourse to the line of reasoning employed in the previous subsection and assume that \textbf{Theory A$'$} and \textbf{Theory B$'$} are probing different regions of the phase diagram controlled by Landau level mixing. In this case, however, we do not have a clear picture of the physical process which might describe an instability of the strongly coupled metal of the $\eta$-fermions to a gapped Laughlin state. 

Let us now consider the inverse scenario in which the $\eta$ fermions fill an integer number of LLs, so that $\nu_\eta = s - 1/2$. Hence,
\begin{align}
	\nu = \frac{s}{s(k-N) + N},  \label{eqn:TheoryB'-IQH-filling}
\end{align}
From Eq. \eqref{eqn:nu-psi}, we see that the $\psi$-fermions are in a metallic state when $1/\nu = k - 1$. This implies
\begin{align}
	\frac{s-1}{s} = N > 0,
\end{align}
for which the only solution is $(N,s)=(2,2)$. In this case, the electronic filling fraction is $	\nu_* = 1/(k-1)$. This simply corresponds to the usual sequence of incompressible states for the fermionic ($k$ odd) and bosonic ($k$ even) Jain sequences. At these fillings, we can integrate out the $\eta$ fermions in \textbf{Theory B$'$} to obtain the effective action
\begin{align}
	\mathcal{L}_{B',\mathrm{eff}} &= \frac{-2}{4\pi} \Tr\left[udu - \frac{2i}{3}u^3\right] + \frac{k-2}{4\pi}bdb - \frac{1}{2\pi} \Tr[u]db + \frac{1}{2\pi}bdA \, , \label{eqn:LB'eff}
\end{align}
describing a non-Abelian topological order. Specifically, this is the Lagrangian for the $[U(2)_{-2}^\mathrm{spin} \times U(1)_{4(k-1)}]/\mathbb{Z}_2 \leftrightarrow [SU(2)_{2} \times U(1)_{4(k-1)}]/\mathbb{Z}_2 = U(2)_{2,2(k-1)}$ Chern-Simons theory. %roughly by $U(2)_{-2} \times U(1)_{k-2}$. \textcolor{violet}{(RS: A quotient is needed to tie the two factors together)}

In order to understand how one arrives at this identification of the Lagrangian as that for a quotient theory, let us start with a decoupled $U(2)_{-2}^\mathrm{spin} \times U(1)_{4(k-1)}$ Chern-Simons theory:
\begin{align}
	\mathcal{L} &= \frac{-2}{4\pi} \Tr\left[\hat{u}d\hat{u} - \frac{2i}{3}\hat{u}^3\right] + \frac{4(k-1)}{4\pi}\hat{b}d\hat{b} + \frac{2}{2\pi}\hat{b}dA \, .
\end{align}
Now, taking the $\mathbb{Z}_2$ quotient of this theory amounts to declaring that $\hat{u}$ and $\hat{b}$ are no longer ``good" gauge fields but the linear combinations
\begin{align}
	u &= \hat{u} - \hat{b}\mathbf{1} \\
	b &= 2\hat{b}
\end{align}
are \cite{Seiberg2016a}. That is to say, we declare $u$ and $b$ to satisfy the appropriate flux quantization conditions. In more formal terms, taking the $\mathbb{Z}_2$ quotient means we gauge the common $\mathbb{Z}_2$ one-form symmetry of the $U(2)_{-2}^\mathrm{spin}$ and $U(1)_{4(k-1)}$ factors \cite{Gaiotto2015,Seiberg2016a} (which is to say, we project out all Wilson lines which have non-trivial braiding with respect to the Wilson line generating the $\mathbb{Z}_2$ one-form symmetry). Rewriting $\mathcal{L}$ in terms of these gauge fields, we arrive at Eq. \eqref{eqn:LB'eff}, as desired.

%Although we do not have a clear physical picture of how this specific topological order would arise in the language of the $\psi$ fermions, that the non-Abelian part is always $SU(2)_{-2}$ is reassuring, as such a topological order could plausibly arise as a paired state of the $\psi$ fermions, which form a metal. %Additionally, the $k=1$ case is troubling, as in this scenario, Theory A describes free Dirac fermions in zero magnetic field, yet Theory B still predicts a non-Abelian topological order. \textcolor{blue}{EF: this is rather vague. First the Dirac fermions are not really free: we just ignored their interactions. I can imagine a scenario in which at weak coupling they are controlled by the free Dirac fixed point. Then, if the interactions are strong enough some quartic coupling may become relevant leading to a paired (Jackiw-Rossi) phase, which is non-Abelian}.

%In this interlude, we illustrated that the considerations applied to the dual fermionic theories appearing in the $SU(2)$ quadrality are relevant to other pairs of dual Abelian and non-Abelian theories. However, although we provided some cursory descriptions of how the states predicted by the dual theories in this subsection may be placed within a consistent physical phase diagram, we do not have \sout{as concrete} a \textcolor{blue}{general} physical picture for the mechanisms underlying the transition between the phases as we were able to provide in the previous subsection. Understanding these instabilities remains an important open problem, which we leave to future work.

\subsection{Examples involving gapped states \label{sec:appendix-UN-duality-gapped-examples}}

Lastly, we can look for filling fractions at which both the $\psi$ and $\eta$ fermions form IQH states. Setting Eqs. \eqref{eqn:TheoryA'-IQH-filling} and \eqref{eqn:TheoryB'-IQH-filling} equal to one another, we find that this happens when
\begin{align}
	N = \frac{s}{s-1} \frac{p}{p-1} %1 = -s\left(\frac{1}{N} + \frac{1}{N(p-1)} - 1 \right).
\end{align}
%In order for the right hand side to equal unity, it is readily checked that $s$ must equal either $1+N(p-1)$ or $1-N(p-1)$. In the former case, the only integer solution for $N$ is $N=1$, which is uninteresting. In the latter case, we arrive at the equation
%\begin{align}
%	p = 1+\frac{2}{N-1}.
%\end{align}
The topological orders predicted by \textbf{Theory A$'$} and \textbf{Theory B$'$} at these filling fractions are described by, respectively, the low energy actions
\begin{align}
	\mathcal{L}_{A',\mathrm{eff}} &= \frac{p-1}{4\pi} ada + \frac{1}{2\pi} adv + \frac{k-1}{4\pi} vdv + \frac{1}{2\pi} vdA \, ,  \\
	\mathcal{L}_{B',\mathrm{eff}} &= -\frac{s}{4\pi} \Tr\left[udu - \frac{2i}{3}u^3\right] + \frac{k-2}{4\pi}bdb - \frac{1}{2\pi} \Tr[u]db + \frac{1}{2\pi}bdA \, .
\end{align}
One integer solution to the above equation is given by $(s,p,N) = (3,4,2)$, corresponding to an electronic filling fraction of $\nu_* = 3 / (3k-4)$.  Here, \textbf{Theory A$'$} predicts a $U(1)^{\mathrm{spin}} \times U(1)$ theory (note $a$ is a $\mathrm{spin}_\mathrm{c}$ connection while $v$ is a regular gauge field) describing the Abelian Jain state at $\nu_* = 2k/(2k-3)$. Using the same quotient construction as in the previous subsection, we can see that \textbf{Theory B$'$} describes a $[U(2)_{-3}^{\mathrm{spin}} \times U(1)_{3(3k-4)}]/\mathbb{Z}_3 \leftrightarrow [SU(3)_{2} \times U(1)_{3(3k-4)}]/\mathbb{Z}_3 = U(3)_{2,3k-4}$ topological order. A second integer solution is given by $(s,p,N) = (4,3,2)$, corresponding to an electronic filling fraction of $\nu_* = 2 / (2k-3)$.  \textbf{Theory A$'$} again predicts an Abelian Jain state, whereas \textbf{Theory B$'$} predicts a non-Abelian $[U(2)_{-4}^{\mathrm{spin}} \times U(1)_{8(2k-3)}]/\mathbb{Z}_4 \leftrightarrow [SU(4)_{2} \times U(1)_{8(2k-3)}]/\mathbb{Z}_4 = U(4)_{2,2(2k-3)}$ topological order.

\end{appendix}

\nocite{apsrev41Control}
\bibliographystyle{apsrev4-1}
\bibliography{pairing}

\end{document}